\pdfoutput=1
\documentclass[11pt,twoside,a4paper,cmspaper,final,collab]{cms-tdr}
\def\svnVersion{cccd813-D}\def\svnDate{2021/08/19}\def\cmsCernNoTag{CERN-EP-2020-251}\def\cmsCernDate{\today}\def\cmsMessage{"Published in the European Physical Journal C as \href{http://dx.doi.org/10.1140/epjc/s10052-021-09570-2}{\doi{10.1140/epjc/s10052-021-09570-2}.}"}
\begin{document}\cmsNoteHeader{SMP-17-008}

\newlength\cmsFigWidth \ifthenelse{\boolean{cms@external}}{\setlength\cmsFigWidth{0.85\columnwidth}}{\setlength\cmsFigWidth{0.4\textwidth}}
\ifthenelse{\boolean{cms@external}}{\newcommand{\cmsLeft}{upper\xspace}}{\newcommand{\cmsLeft}{left\xspace}}
\ifthenelse{\boolean{cms@external}}{\newcommand{\cmsRight}{lower\xspace}}{\newcommand{\cmsRight}{right\xspace}}
\newlength\cmsTabSkip\setlength{\cmsTabSkip}{1ex}
\newcommand{\cmsTable}[1]{\resizebox{\textwidth}{!}{#1}}

\newcommand{\deltaR}{\ensuremath{\Delta R_{23}\xspace}} 
\newcommand{\lops}{LO 2j+PS}
\newcommand{\nlops}{NLO 2j+PS} 
\newcommand{\lofourps}{LO 4j+PS} 
\newcommand{\Zjet}{\PZ + two-jet}
\newcommand{\jetratio}{\ensuremath{p_{\mathrm{T3}}\xspace/p_{\mathrm{T2}}\xspace}} 
\newcommand{\ptl}{\ensuremath{p_{\mathrm{T1}}}\xspace}
\newcommand{\ptnl}{\ensuremath{p_{\mathrm{T2}}}\xspace} 
\newcommand{\ptnnl}{\ensuremath{p_{\mathrm{T3}}}\xspace}
\newcommand{\ptz}{\ensuremath{p_{\mathrm{T}\PZ}}\xspace} 
\newcommand{\RooUnfold} {{\textsc{RooUnfold}}\xspace}

\renewcommand{\labelenumi}{(\roman{enumi})}

\cmsNoteHeader{SMP-17-008}

\title{Measurements of angular distance and momentum ratio distributions in three-jet and \texorpdfstring{$\PZ$}{Z} + two-jet final states in \texorpdfstring{$\Pp\Pp$}{pp} collisions}
\titlerunning{Measurements of three-jet and Z + two-jet distributions}
\date{\today}

\abstract{ Collinear (small-angle) and large-angle, as well as soft and hard radiations are investigated in three-jet and \PZ + two-jet events collected in proton-proton collisions at the LHC.
The normalized production cross sections are measured as a function of the ratio of transverse momenta of two jets and their angular separation.
The measurements in the three-jet and \PZ + two-jet events are based on data collected at a center-of-mass energy of 8\TeV, corresponding to an integrated luminosity of 19.8\fbinv.
The \PZ + two-jet events are reconstructed in the dimuon decay channel of the \PZ boson.
The three-jet measurement is extended to include $\sqrt{s} = 13\TeV$ data corresponding to an integrated luminosity of 2.3\fbinv.
The results are compared to predictions from event generators that include parton showers, multiple parton interactions, and hadronization.
The collinear and soft regions are in general well described by parton showers, whereas the regions of large angular separation are often best described by calculations using higher-order matrix elements. }

\hypersetup{
pdfauthor={CMS Collaboration},
pdftitle={Measurements of angular distance and momentum ratio distributions in three-jet and Z + two-jet final states in pp collisions},
pdfsubject={CMS},
pdfkeywords={CMS,  QCD, jet}
}

\maketitle

\section{Introduction}

Collimated streams of particles, produced in interactions of quarks and gluons and reconstructed as jets, are described by the theory of strong interactions, quantum chromodynamics (QCD).
Multijet events provide exemplary signatures in high-energy collider experiments, and modeling their characteristics plays an important role in precision measurements, as well as in searches for new physics.
The understanding of the structure of multijet final states is therefore crucial for analyses of those events.

Theoretical predictions for multijet events are based on a matrix element (ME) expansion to a fixed perturbative order, supplemented by the parton shower (PS) approach to approximate higher-order perturbative contributions.
The ME expansion incorporates color correlations between quarks and gluons, including interference terms, as well as kinematic correlations between the partons, without any approximation at fixed perturbative order.
Its application is, however, currently limited to final states with less than $\mathcal{O}(10)$ partons.
The PS can simulate final states containing many partons, but with probabilities calculated using the approximations of soft and collinear kinematics and partial or averaged color structures.
The best descriptions of multijet final states are based on a combination of both approaches~\cite{Catani:2001cc,Buckley:2011ms,Bengtsson:1986hr,Mrenna:2003if}.
Other features implemented in simulations, such as multiple parton interactions (MPI) and hadronization, also play an important role, \eg, in describing angular correlations between jets~\cite{Chatrchyan:2013fha,Abe:1994nj,Abbott:1997bk}.

\begin{figure*}[ht] 
\centering
\includegraphics[width=0.7\textwidth]{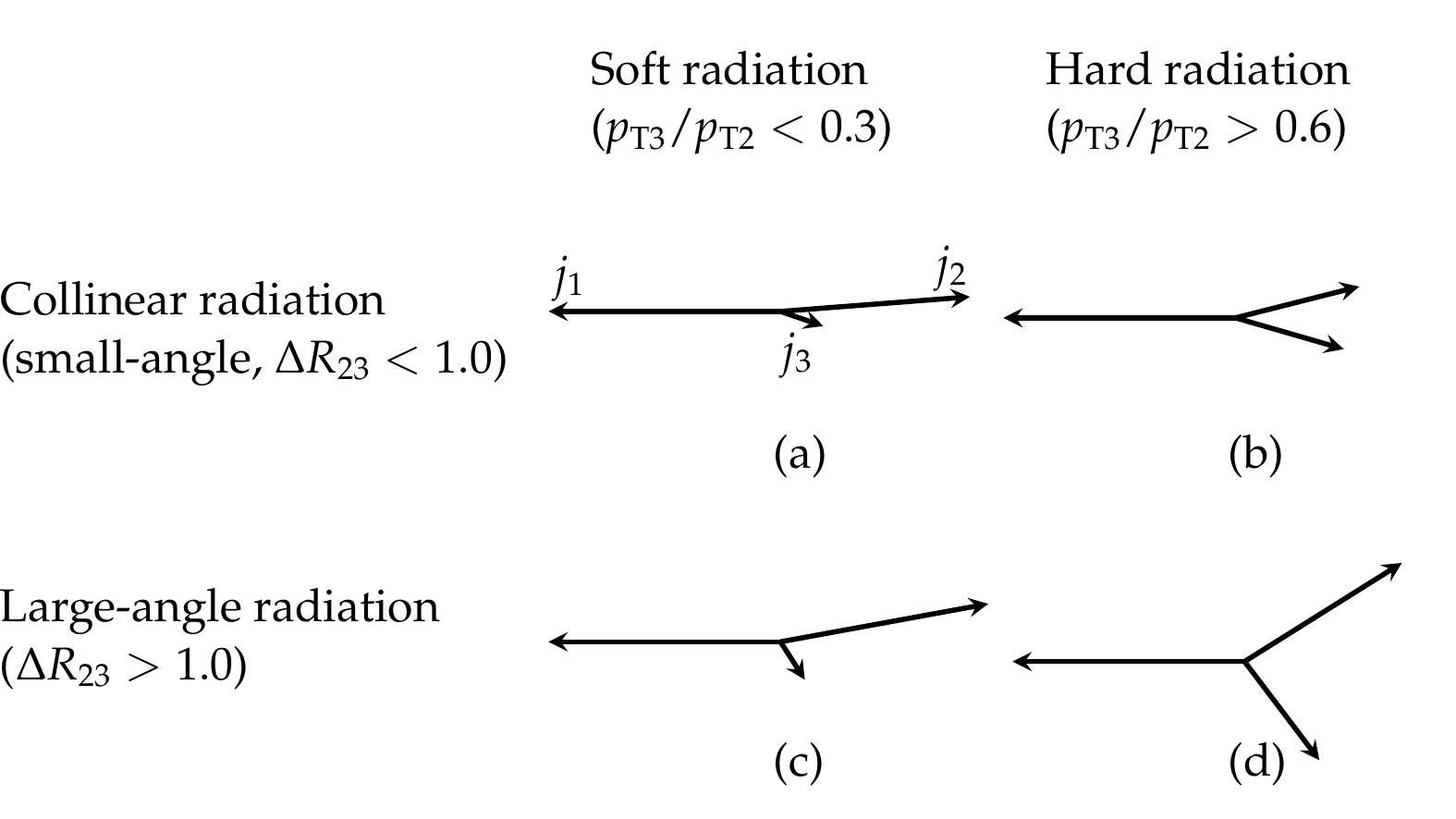} 
\caption{\label{fig1} Four categories of parton radiation. (a)~soft and small-angle radiation, (b)~hard and small-angle radiation, (c)~soft and large-angle radiation, (d)~hard and large-angle radiation.} 
\end{figure*}

In this paper, we investigate collinear (small-angle) and large-angle radiation in different regions of jet transverse momentum (\pt) by concentrating on two different topologies, one using three-jet events and another with \Zjet\ events.
We label the hardest jet, or \PZ boson as $j_1$, the next hardest as $j_2$, and the softest as $j_3$.
We introduce two observables that are sensitive to the dynamic properties of multijet final states.
One observable is the \pt ratio of $j_3$ to $j_2$, \jetratio.
The other observable is the angular distance between the jet centers of $j_2$ and $j_3$ in the rapidity-azimuth ($y$-$\phi$) phase space, $\deltaR = \sqrt{\smash[b]{(y_{3} -y_{2})^{2} + (\phi_{3} - \phi_{2})^{2}}}$.
The definition of rapidity is $y = \ln\sqrt{(E+p_{z}c)/(E-p_{z}c)}$, and the definitions of other kinematic variables are given in Ref.~\cite{Chatrchyan:2008zzk}.
As indicated in Fig.~\ref{fig1}, we classify three-jet and \Zjet\ events into different categories using these two observables:
\begin{enumerate} 
\item soft ($\jetratio < 0.3$) or hard ($\jetratio > 0.6$) radiation, depending on the ratio \protect\jetratio; 
\item small-angle ($\deltaR < 1.0$) or large-angle ($\deltaR > 1.0$) radiation, depending on the angular separation \deltaR. 
\end{enumerate}
According to these classifications, events in the soft and small-angle radiation region, as shown in Fig.~\ref{fig1} (a), can only be described if soft gluon resummation, \eg, in form of a parton shower, is included, whereas events in the hard and large-angle radiation region, as shown in Fig.~\ref{fig1} (d), would be better described when including the ME calculations.
The events in Figs.~\ref{fig1} (b) and (c) are also of interest, since they should include effects from both the PS and ME.

We report on proton-proton ($\Pp\Pp$) collision data collected at the CMS experiment containing three-jet events at center-of-mass energies of 8 and 13\TeV, and \Zjet\ events at a center-of-mass energy of 8\TeV.
The measurements are compared to calculations based on a leading-order (LO) or next-to-leading-order (NLO) ME supplemented with effects from PS, MPI, and hadronization.
The NLO ME descriptions apply to the lowest parton multiplicities relevant to the selected events: 2 jets for the three-jet analysis and \PZ+1j for the \Zjet\ analysis.
The measurements using three-jet final states are complementary to those with \Zjet\ events in a sense that different kinematic regions and initial-state flavor compositions are being probed.
The jets are also fully color connected, while the \PZ boson is color neutral, so color coherence effects should not appear so strongly in \Zjet\ events. 

The goal of the measurements is: (i) to untangle the different features of the radiation in the collinear and large-angle events; 
(ii) to investigate how well the PS approach describes the hard and large-angle radiation patterns;
and (iii) to illustrate how ME calculations can attempt to describe the soft and collinear regions.

\section{The CMS detector}

{\tolerance=800
The central feature of the CMS detector is a superconducting solenoid of 6\unit{m} internal diameter, providing a magnetic field of 3.8\unit{T}. 
A silicon pixel and strip tracker, a lead tungstate crystal electromagnetic calorimeter (ECAL), and a brass and scintillator hadron calorimeter (HCAL), each composed of a barrel and two endcap sections, reside within the volume of the solenoid.
Charged-particle trajectories are measured in the tracker with full azimuthal acceptance within pseudorapidities $\abs{\eta} < 2.5$. 
The ECAL, which is equipped with a preshower detector in the endcaps, and the HCAL cover the region $\abs{\eta} < 3.0$.
Forward calorimeters extend the pseudorapidity coverage provided by the barrel and endcap detectors to the region $3.0<\abs{\eta} < 5.2$.
Finally, muons are measured up to $\abs{\eta} < 2.4$ in gas-ionization detectors embedded in the steel flux-return yoke outside the solenoid.
Events of interest are selected using a two-tiered trigger system~\cite{Khachatryan:2016bia}.
The first level, composed of custom hardware processors, uses information from the calorimeters and muon detectors to select events at a rate of around 100\unit{kHz} within a fixed latency of about 4\mus.
The second level, known as the high-level trigger (HLT), consists of a farm of processors running a version of the full event reconstruction software optimized for fast processing, and reduces the event rate to around 1\unit{kHz} before data storage.
\par}

A more detailed description of the CMS detector, together with a definition of the coordinate system and the kinematic variables, is given in Ref.~\cite{Chatrchyan:2008zzk}.

\section{Event samples and selection}

The data in this study were collected with the CMS detector at the LHC using pp collisions at center-of-mass energies of 8 and 13\TeV.
The $\sqrt{s} = 8\TeV$ data, taken in 2012 during LHC Run 1, correspond to an integrated luminosity of 19.8\fbinv, and the $\sqrt{s} = 13\TeV$ data, taken in 2015 during LHC Run 2, correspond to an integrated luminosity of 2.3\fbinv.

Particles are reconstructed and identified using a particle-flow (PF) algorithm~\cite{bib:parflow}, that utilizes an optimized combination of information from the various elements of the CMS detector.
Jets are reconstructed by clustering the four-vectors of the PF candidates with the infrared and collinear-safe anti-\kt clustering algorithm~\cite{Cacciari:2008gp} using a distance parameter $R_{\mathrm{jet}} =$ 0.5 (0.4) at $\sqrt{s} =  8\,(13)\TeV$.
The clustering is performed with the \FASTJET software package~\cite{Cacciari:2011ma}.
The jets are ordered in \pt and all events with additional jets are analyzed.
In addition, three-jet events use the charged-hadron subtraction (CHS) technique~\cite{bib:parflow} to mitigate the effect of extraneous $\Pp\Pp$ collisions in the same bunch crossing (pileup, PU).
The CHS technique reduces the contribution to the reconstructed jets from PU by removing tracks identified as originating from PU vertices.

Muons are reconstructed using a simultaneous global fit performed with the hits in the silicon tracker and the muon system. 
They are required to pass standard identification criteria~\cite{Chatrchyan:2013sba,Sirunyan:2018fpa} based on the minimum number of hits in each detector, quality of the fit, and the consistency with the primary vertex by requiring the longitudinal (transverse) impact parameters to be less than 0.5 (0.2) \cm. 
The efficiency to reconstruct and identify muons is greater than 95\% over the entire region of pseudorapidity covered by the CMS muon system ($\abs{\eta} > 2.4$).
The overall momentum scale is measured to a precision of 0.2\% with muons from \PZ decays.
The transverse momentum resolution varies from 1\% to 6\% depending on pseudorapidity for muons with \pt for a few \GeV to 100\GeV and reaches 10\% for 1\TeV muons~\cite{Chatrchyan:2012xi}.
Observed distributions for muons are well reproduced by Monte Carlo (MC) simulation. 
Corresponding scale factors for the difference between data and MC simulations are measured with good accuracy~\cite{CMS-DP-2013-009}.
Muons must be isolated from other activity in the tracker by requiring the \pt sum of other tracks within a cone of radius $\Delta R = \sqrt{(\Delta\eta)^{2} + (\Delta\phi)^{2}} = 0.3$ centered on the muon candidate, is less than 10\% of the muon \pt.
If the two muons with the highest \pt in an event are within the isolation cone of one another, the other muon candidate is removed from the isolation sum of each muon.

{\tolerance=1600
Three-jet events are collected using single jet HLT requirements that are not pre-scaled.
The $\sqrt{s} =  8\,(13)\TeV$ data use a 320 (450)\GeV trigger \pt threshold. 
In the offline analyses, the \pt threshold starts at 510\GeV for both sets of data.
The \Zjet\ events with the \PZ boson decaying into a pair of muons are collected at $\sqrt{s} = 8\TeV$ with a single-muon HLT that requires a muon $\pt > 24\GeV$ and $\abs{\eta} < 2.1$.
\par}

In the three-jet systems, the leading jet is required to have a $\pt > 510\GeV$, because of a decreasing efficiency for single jet triggers below this value~\cite{Khachatryan:2016bia, Khachatryan:2016mlc, Khachatryan:2016wdh}.
Events with at least three jets of $\pt > 30\GeV$ are selected for further consideration.
The leading and subleading jets must be within a rapidity range of $\abs{y} < 2.5$, and the third jet is therefore implicitly restricted to $\abs{y} < 4$ by requiring $\deltaR < 1.5$.
A dijet topology with an extra jet is selected by requiring the difference in azimuthal angle between the first and second jet to be $ \pi-1 < \Delta \phi_{12} < \pi$.
The missing transverse momentum vector \ptvecmiss is defined as the projection onto the plane perpendicular to the beam axis of the negative vector sum of the momentum of all reconstructed PF objects in an event.
Its magnitude is referred to as \ptmiss.
Events with a \ptmiss divided by the scalar sum of all transverse momenta $> 0.3$ are rejected to remove the contamination from \PW or \PZ boson decays~\cite{CMS-PAPERS-JME-10-009, CMS-PAPERS-JME-13-003, CMS-PAPERS-JME-17-001}.
To avoid an overlap between $j_2$ and $j_3$, \deltaR\ is required to be larger than the distance parameter $R_{\mathrm{jet}}$.
We thus require \deltaR\ to be larger than 0.6 (0.5) for $\sqrt{s} =  8\,(13)\TeV$ data. The maximum \deltaR\ is set to 1.5 to ensure that $j_3$ is closer to $j_2$ than to $j_1$.
We further require that $0.1<\jetratio\ < 0.9$ to avoid \ptnnl threshold effects and to ensure \pt ordering for hard radiation.

In \Zjet\ events, the \PZ boson is reconstructed from a pair of oppositely charged, isolated muons with $\pt > 25~(5)\GeV$ and $\abs{y} < 2.1$ (2.4) for the leading (subleading) muon.
Muons are required to be from primary vertex with distance $dr < 0.2 \cm$ and $dz < 0.5 \cm$.
The dimuon invariant mass is required to be $70 < m_{\mu^+\mu^-} < 110\GeV$ with the dimuon momentum satisfying $\ptl > 80\GeV$ and $\abs{y_1} < 2$.
At least two jets are required in the final state with the leading jet (labeled $j_2$) satisfying $\ptnl > 80\GeV$ and $\abs{y_{2}} < 1$ and the subleading jet (labeled $j_3$) required to have $\ptnnl > 20\GeV$ with $\abs{y_{3}} < 2.4$.
The distance between muons from \PZ bosons and jets are requested to be more then 0.5.
The \Zjet\ topology is further restricted by requiring a difference in the azimuthal angle between the \PZ boson and  $j_{2}$ of $ \Delta\phi_{12} > 2$. 

Table~\ref{tabphasespace} shows a summary of the event selection requirements for both samples.

\begin{table*}[htbp] 
\centering 
\topcaption{Phase space selection for the three-jet and \Zjet\ analyses.} 
\label{tabphasespace}
\begin{tabular}{ l l } 
Three-jet events &  \\ 
\hline 
Transverse momentum of the leading jet ($j_1$) & $\ptl > 510\GeV$\\ 
Transverse momentum of each jet and rapidity of $j_{1,2}$ & $\pt > 30\GeV$ , $\abs{y_{1,2}} < 2.5 $\\ 
Azimuthal angle difference between $j_1$ and $j_2$ & $\pi-1 < \Delta\phi_{12} < \pi$ \\ 
Transverse momentum ratio between $j_2$ and $j_3$ & $0.1 <\jetratio < 0.9$ \\ 
Angular distance between $j_2$ and $j_3$ & $R_{\mathrm{jet}}+0.1 < \deltaR < 1.5$ \\
Number of selected events at $\sqrt{s} =  8\,(13) \TeV$ & 777\,618 (613\,254) \\
\\[-1.5ex]
\Zjet\ events &  \\ 
\hline 
Transverse momentum of the \PZ boson ($j_1$) & $\ptl > 80\GeV$, $\abs{y_1} < 2$ \\
Transverse momentum and rapidity of $j_2$ & $\ptnl > 80 \GeV$ , $\abs{y_{2}} < 1 $\\
Transverse momentum and rapidity of $j_3$ & $\ptnnl > 20 \GeV $, $\abs{y_{3}} < 2.4 $\\
Azimuthal angle difference between \PZ and  $j_2$ & $ 2 < \abs{\Delta \phi_{12}} < \pi $ \\
Dimuon mass & $70< m_{\mu^+\mu^-} < 110 \GeV$ \\ 
Angular distance between $j_3$ and $j_2$ & $0.5 < \deltaR < 1.5$ \\ 
Number of selected events & 15\,466\\
\end{tabular} 
\end{table*}

Generator jets are reconstructed from stable particles by clustering the four-vectors with an anti-\kt clustering algorithm excluding neutrinos. 
The kinematical rerquirements for muons and jets are the same as applied for reconstructed objects. 
For \Zjet\ events, the distance between muons from \PZ boson and jets must have $\Delta R > 0.5$. 
The \ptmiss selection is not applied at the generator level for QCD multijet events.

\section{Theoretical models} \label{sec:theory}

{\tolerance=1600
Reconstructed data are compared to predictions from MC event generators, where the generated events are passed through a full detector simulation based on \GEANTfour~\cite{bib:geant} and the simulated events are reconstructed using standard CMS software.
Reconstruction-level predictions are obtained for three-jet events at $\sqrt{s}= 8\TeV$ with the \MADGRAPH~\cite{bib:madgraph5} software package matched to \PYTHIA~6~\cite{Sjostrand:2006za} with the CTEQ6L1~\cite{Pumplin:2002vw} parton distribution function (PDF) set and the Z2Star tune~\cite{CMS-PAPERS-QCD-10-010}, as well as with standalone \PYTHIA~8.1~\cite{Sjostrand:2007gs} with the CTEQ6L1 PDF set and the 4C~\cite{Corke_2011} tune.
At 13\TeV, \MADGRAPH interfaced to \PYTHIA~8.2~\cite{Sjostrand:2014zea} and standalone \PYTHIA~8.2 are used with the NNPDF2.3LO~\cite{Ball:2012cx} PDF set and the  CUETP8M1~\cite{Khachatryan:2015pea} tune.
The \SHERPA~\cite{Gleisberg:2008ta} event generator interfaced to {\textsc{csshower++}}~\cite{Schumann:2007mg} with the CT10~\cite{Lai:2010vv} PDF set and the AMISIC++~\cite{PhysRevD.36.2019} tune and \MADGRAPH interfaced to \PYTHIA~6 with the CTEQ6L1 PDF set and the Z2Star tune provide \Zjet\ events at 8\TeV.
Table~\ref{detMC} summarizes the event generator versions, PDF sets and tunes.
\par}

\begin{table*}[htbp] 
\centering 
\topcaption{Event generator versions, PDF sets, and tunes used to produce MC samples at reconstruction level.} 
\label{detMC} 
\begin{tabular}{l l l} 
Event generator & PDF set & Tune \\
Three-jet events at $\sqrt{s} = 8\TeV$ & & \\
\hline
\MADGRAPH~5.1.3.30 + \PYTHIA~6.425 & CTEQ6L1 & Z2Star \\ 
\PYTHIA~8.153  & CTEQ6L1 & 4C \\ 
\\[-1.5ex]
 Three-jet events at $\sqrt{s} = 13\TeV$ & & \\
\hline 
\MADGRAPH~5.2.3.3 + \PYTHIA~8.219 & NNPDF2.3LO & CUETP8M1 \\ 
\PYTHIA~8.219 & NNPDF2.3LO & CUETP8M1 \\ 
\\[-1.5ex]
\Zjet\ events & & \\ 
\hline 
\SHERPA~1.4.0 + {\textsc{csshower++}} & CT10 & AMISIC++  \\  
\MADGRAPH~5.1.3.30 + \PYTHIA~6.425 & CTEQ6L1  & Z2Star \\ 
\end{tabular} 
\end{table*}

Results corrected to stable-particle level are compared to predictions obtained with the models presented below.
An overview of these models is given in Table~\ref{tableMC}.

The \PYTHIA~8~\cite{Sjostrand:2014zea} event generator provides hard-scattering events using a ME calculated at LO supplemented with PS.
These event samples are labeled as ``\PYTHIA \lops" for the three-jet and as ``\PYTHIA LO Z+1j+PS" for \Zjet\ events.
The PDF set NNPDF2.3LO and the CUETP8M1 parameter set for the simulation of the underlying event (UE) are used with free parameters adjusted to measurements in $\Pp\Pp$ collisions at the LHC and proton-antiproton collisions at the Fermilab Tevatron.
The Lund string model~\cite{Andersson:1998tv} is applied for the hadronization process.

The \MGvATNLO event generator, labeled as ``\MADGRAPH{}" in the following, is used to simulate hard processes with up to 4 final-state partons at LO accuracy.
It is interfaced to \PYTHIA~8 with the CUETP8M1 tune and the NNPDF2.3LO PDF set for the simulation of PS, hadronization, and MPI, for three-jet, and to \PYTHIA~6 with the Z2Star tune and the CTEQ6L1 PDF set for \Zjet\ events.
The three-jet sample is labeled as ``\MADGRAPH \lofourps" and the \Zjet\ sample is labeled as ``\MADGRAPH LO Z+4j+PS".
The \kt-MLM procedure \cite{Alwall:2007fs} is used to match jets from the ME and PS with a matching scale of 10\GeV. 

Predictions are also included using the \POWHEG\ {\textsc{box}} library \cite{bib:Nason:2004rx,bib:Frixione:2007vw,bib:Alioli:2010xd}, with the CT10 NLO~\cite{Lai:2010vv} PDFs and with the \PYTHIA~8 CUETP8M1 tune applied to simulate PS, MPI, and hadronization.
The \POWHEG generator is run in the dijet mode \cite{bib:POWHEG_Dijet} providing an NLO $2\to2$ calculation, labeled as ``\POWHEG \nlops".
The matching between the \POWHEG ME calculations and the \PYTHIA UE~\cite{Khachatryan:2015pea} simulation is performed using the shower-veto procedure (UserHook option 2~\cite{Sjostrand:2014zea}).

The \SHERPA software package is used to simulate \Zjet\ events.
The hard process is calculated at LO for a ME with up to four final-state partons and the CT10 PDF set is used.
This sample is labeled as ``\SHERPA LO Z+4j+PS". The \SHERPA\ generator has its own PS~\cite{Schumann:2007mg}, hadronization, and MPI tune~\cite{PhysRevD.36.2019}.

Finally, the \MGvATNLO generator is also used in the \MCATNLO mode, providing a \PZ + one-jet ME at NLO accuracy.
This event generator is interfaced to \PYTHIA~8, using the CUETP8M1 tune and the NNPDF3.0NLO~\cite{Ball:2014uwa} PDF set, to produce \Zjet\ events. The sample is labeled as ``a\MCATNLO NLO Z+1j+PS". 

The background from \PW, \PZ, top quark, and diboson production for the three-jet analysis is negligible and not further considered.
The main background for \Zjet\ events comes from \ttbar, single top, and diboson production.
The \ttbar, ZZ, and WZ events are simulated with \MADGRAPH 5.1.3.30 + \PYTHIA 6.425 using the same tune and PDF set as for generating \Zjet\ samples.
WW events are generated with \PYTHIA 6.425 with CTEQ6L1 PDF set and Z2Star tune.
Single top events are generated with \POWHEG (CT10 PDF set, Z2Star tune).

\begin{table*}[htbp] 
\centering 
\topcaption{MC event generators and version numbers, parton-level processes, PDF sets, and UE tunes used for the comparison with measurements.} 
\label{tableMC} 
\cmsTable{ 
\begin{tabular}{l l l l} 
Event generator & {Parton-level process} & {PDF set} & {Tune}\\ 
\\[-1.5ex]
{Three-jet events} & & & \\
\hline \PYTHIA~8.219 & \lops & NNPDF2.3LO & CUETP8M1 \\ 
\MADGRAPH~5.2.3.3 + \PYTHIA~8.219 & \lofourps & NNPDF2.3LO & CUETP8M1 \\ 
\POWHEG 2 + \PYTHIA~8.219 & \nlops & CT10 NLO & CUETP8M1 \\ 
\\[-1.5ex]
{\Zjet\ events} & & &  \\ 
\hline 
\PYTHIA~8.219 & LO Z+1j+PS & NNPDF2.3LO & CUETP8M1 \\ 
\MADGRAPH~5.1.3.30 + \PYTHIA~6.425 & LO Z+4j+PS & CTEQ6L1 & Z2Star \\ 
\SHERPA~1.4.0 + {\textsc{csshower++}} & LO Z+4j+PS & CT10 & AMISIC++ \\ 
a\MCATNLO + \PYTHIA~8.223 & NLO Z+1j+PS & NNPDF30\_nlo\_nf\_5\_pdfas & CUETP8M1 \\ 
\\[-1.5ex]
\end{tabular} }  
\end{table*}

\section{Data correction and study of systematic uncertainties} \label{sec:systematic}

To facilitate the comparison of data with theory, the data are unfolded from reconstruction to stable-particle level, defined by a mean decay length larger than 1\unit{cm}, so that measurement effects are removed and that the true distributions in the observables are determined.
The unfolding is performed using the D'Agostini algorithm~\cite{DAgostini:1994zf} as implemented in the \RooUnfold\ software package~\cite{bib:RooUnfold} for three-jet events, while the singular value decomposition method~\cite{Hocker:1995kb} is used for \Zjet\ events.
The response matrices are obtained from the full detector simulation using \MADGRAPH for three-jet events and \SHERPA\ for  \Zjet\ events.

We estimate the influence of \ttbar, single top, and diboson backgrounds by adding generated events produced with event generator \MADGRAPH LO Z+4j+PS and comparing the predictions for the observables \jetratio\ and \deltaR\ using the same generator without the backgrounds.
For \ttbar production with fully leptonic decay and dibosons the probability of $j_3$ emission increases from 2\% (soft radiation) to 10\% (hard radiation) depending on the phase space.
For semileptonic and hadronic decays and single top production the change is negligible.
Since the background effect is comparable to the systematic uncertainties, it is not included in the theoretical estimations and it is not subtracted from the data.

The distributions are normalized to the integral of the spectra for three-jet events and to the number of inclusive $\PZ$ + one-jet events in the \Zjet\ analysis.
The \Zjet\ analysis normalization thus reflects the probability to have more than one jet in the event.

Systematic uncertainties associated to the jet energy scale (JES) calibration, the jet energy resolution (JER), PU modeling, model dependence, as well as the unfolding method, are estimated.
Muon-related uncertainties (single muon trigger efficiency, muon isolation, muon scale and resolution) for the \Zjet\ channel are negligible with respect to other systematic sources.
The treatment of the uncertainty depends on the uncertainty source and is estimated separately for each bin (see below).
The overall uncertainty for each bin is estimated summing in quadrature uncertainties from the various sources.

The systematic uncertainty from the JES is 0.15 (0.24)\% at $\sqrt{s} =  8\,(13)\TeV$ for the three-jet case and 5--10\% for the \Zjet\ events.
The JER observed in data differs from that obtained from simulation and simulated jets are therefore smeared to obtain the same resolution as in the data~\cite{bib:jes8}.
The systematic uncertainty from JER is estimated by varying the simulated JER uncertainty up and down by one standard deviation, which results in a systematic uncertainty of 0.16 (0.12)\% at $\sqrt{s} =  8\,(13)\TeV$ for three-jet and 2--3\% for \Zjet\ events.
When the distributions of \Zjet\ events are normalized to the integrals of the histograms, instead of the number of \PZ + one-jet events, the systematic uncertainties due to the JES and JER decrease to 0.3--0.5\%, except for the \jetratio\ shape, which is still sensitive to the JES with changes of up to 3\%.

The distribution in the number of primary vertices is sensitive to the PU difference between data and simulation.
To estimate the uncertainty due to the PU modeling, the number of PU events in simulation is changed by shifting the total inelastic cross section by $\pm$5\%~\cite{Chatrchyan:2012nj}.
The resulting PU uncertainties are 0.10 (0.17)\% at $\sqrt{s} = 8\,(13)\TeV$ for the three-jet and 1\% for the \Zjet\ events.

The dependence on the event generator used for the unfolding is estimated with MC event samples from \MADGRAPH and \PYTHIA for three-jet, and \SHERPA\ and \MADGRAPH\ for the \Zjet\ events.
The means of both sets of unfolded data are used as the nominal values.
This uncertainty is $\approx 1.1$ (0.25)\% at $\sqrt{s} =  8\,(13)\TeV$ for the three-jet and 1\% for the \Zjet\ events, which is half of the difference between the results obtained with the respective event generators.
The difference in the results is due to statistical fluctuations from the limited number of events in the MC simulation.

Table~\ref{sys} summarizes the systematic uncertainties in the measurements.

\begin{table*}[htbp] 
\centering 
\topcaption{Systematic uncertainties in the measurements in \%.} 
\label{sys} 
\begin{tabular}{l c c} 
{Source} & three-jet 8/13\TeV& {\Zjet\ 8\TeV} \\ 
\hline 
Jet energy scale     & 0.15/0.24  & 5--10 \\ 
Jet energy resolution  & 0.16/0.12  & 2--3 \\ 
Pileup              & 0.1/0.17  & 1 \\ 
Unfolding and model dependence   & 1.1/0.25 & 1\\
\end{tabular} 
\end{table*}

The systematic uncertainties from various sources are similar for the three-jet samples at $\sqrt{s} = 8$ and 13 \TeV, except for unfolding and model dependence at $\sqrt{s} = 8\TeV$.
The systematic uncertainties between the three-jet and \Zjet\ analysis cannot be compared directly because each analysis uses a different normalization and also differs in statistical significance.
The JES uncertainty is especially sensitive to the jet \pt range, and the \Zjet\ phase space has a lower \pt threshold than the one used in the three-jet events.

The figures of Sec.~\ref{sec:results} show the total systematic uncertainty as a band in the panels displaying the ratio of predictions over data.

\section{Results} \label{sec:results}

We compare the distributions in the ratio \jetratio\ in data to predictions for events with small-angle  ($\deltaR < 1.0$) and large-angle radiation ($\deltaR > 1.0$).
We also compare the \deltaR\ distributions in data to predictions with soft ($\jetratio < 0.3$) and  hard radiation ($\jetratio > 0.6$).
The events with $0.3 < \jetratio < 0.6$ are not used in the comparisons for the \deltaR\ observable because we focus on the limits in soft and hard radiation.
This classification is summarized in Fig.~\ref{fig1}, within the phase space defined in Table~\ref{tabphasespace}. 
The data measurements are provided at the Durham High Energy Physics Database (HEPData)~\cite{hepdata}.

The uncertainties in the PDF and in the renormalization and factorization scales are investigated for the \POWHEG and a\MCATNLO\ models. 
Other theoretical predictions are expected to have comparable uncertainties.
The PDF uncertainties are calculated as recommended in PDF4LHC~\cite{Butterworth:2015oua} following the description of the PDF sets: for CT10 using the Hessian approach; and for NNPDF using MC replicas.
The renormalization and factorization scales are varied by a factor 2 up and down, excluding the (2,1/2) and (1/2,2) cases.
Finally, the theoretical uncertainties are obtained as the quadratic sum of the PDF variance and the envelope of the scale variations, and displayed as a band around the theoretical predictions in the Fig.~\ref{fig:8TeVpt}--\ref{fig:ZfigDelta}.

\subsection{Three-jet selection}

\begin{figure}[htb] 
\centering 
\includegraphics[width=0.4\textwidth]{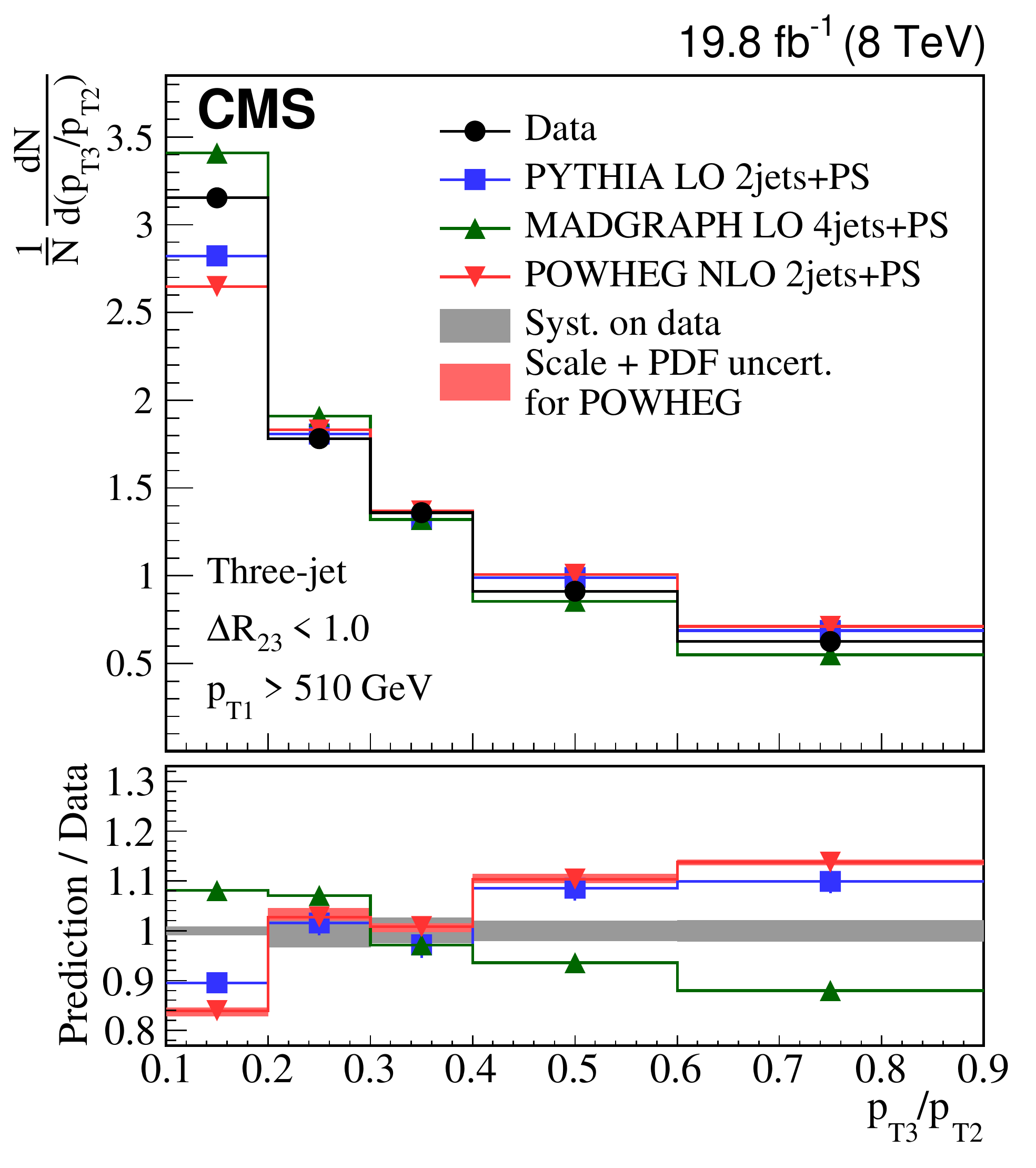} 
\includegraphics[width=0.4\textwidth]{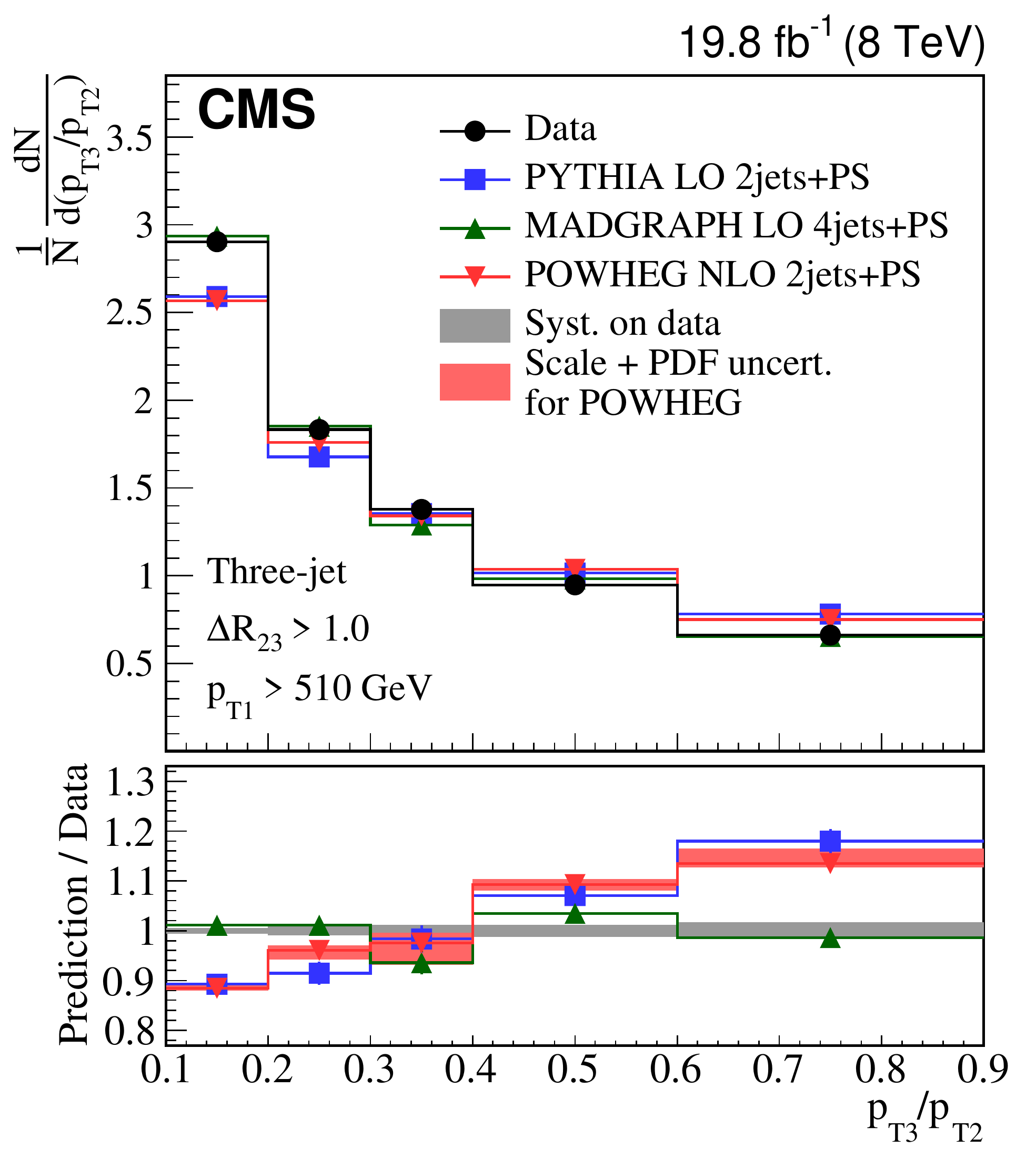} 
\caption{Three-jet events at $\sqrt{s} = 8\TeV$ compared to theory: (\cmsLeft) \protect\jetratio\ for small-angle radiation ($\deltaR < 1.0$), (\cmsRight) \protect\jetratio\ for large-angle radiation ($\deltaR > 1.0$).} 
\label{fig:8TeVpt} 
\end{figure}

\begin{figure}[htb] 
\centering 
\includegraphics[width=0.4\textwidth]{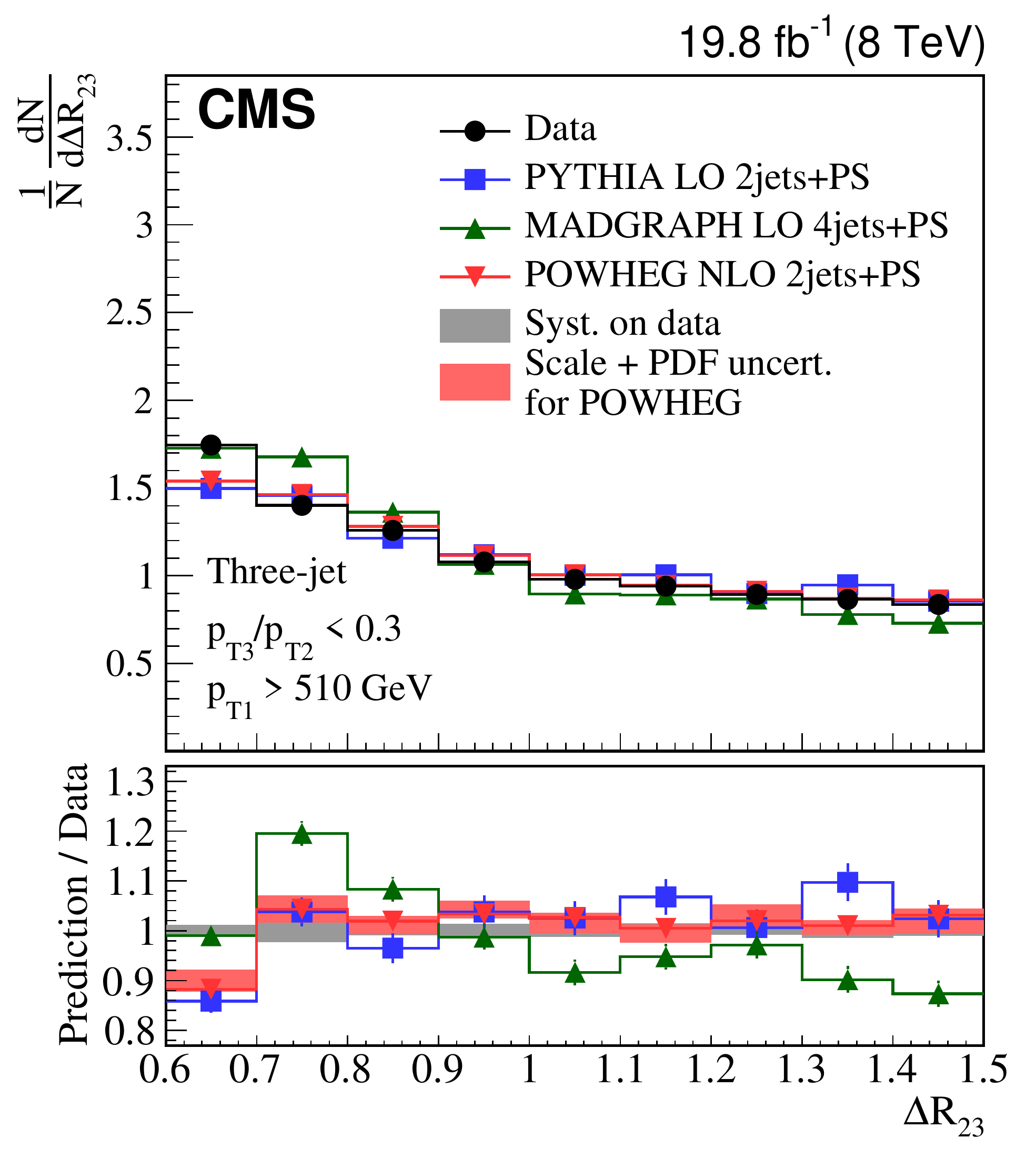} 
\includegraphics[width=0.4\textwidth]{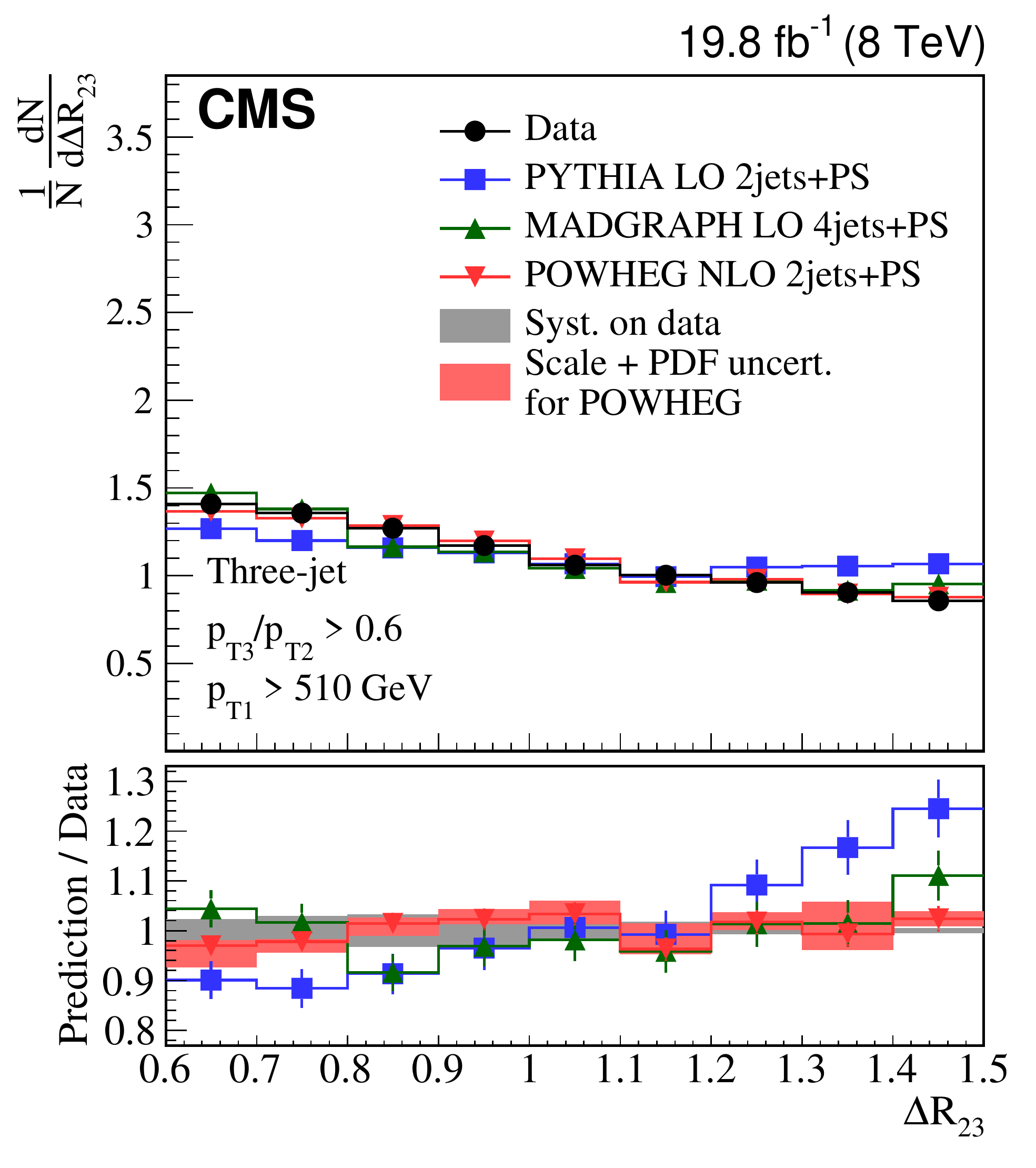} 
\caption{Three-jet events at $\sqrt{s} = 8\TeV$ and comparison to theoretical predictions: (\cmsLeft) \deltaR\ for soft radiation ($\jetratio < 0.3$), (\cmsRight) \deltaR\ for hard radiation ($\jetratio > 0.6$).} 
\label{fig:8TeVDelta}
\end{figure}

\begin{figure}[htb] 
\centering 
\includegraphics[width=0.4\textwidth]{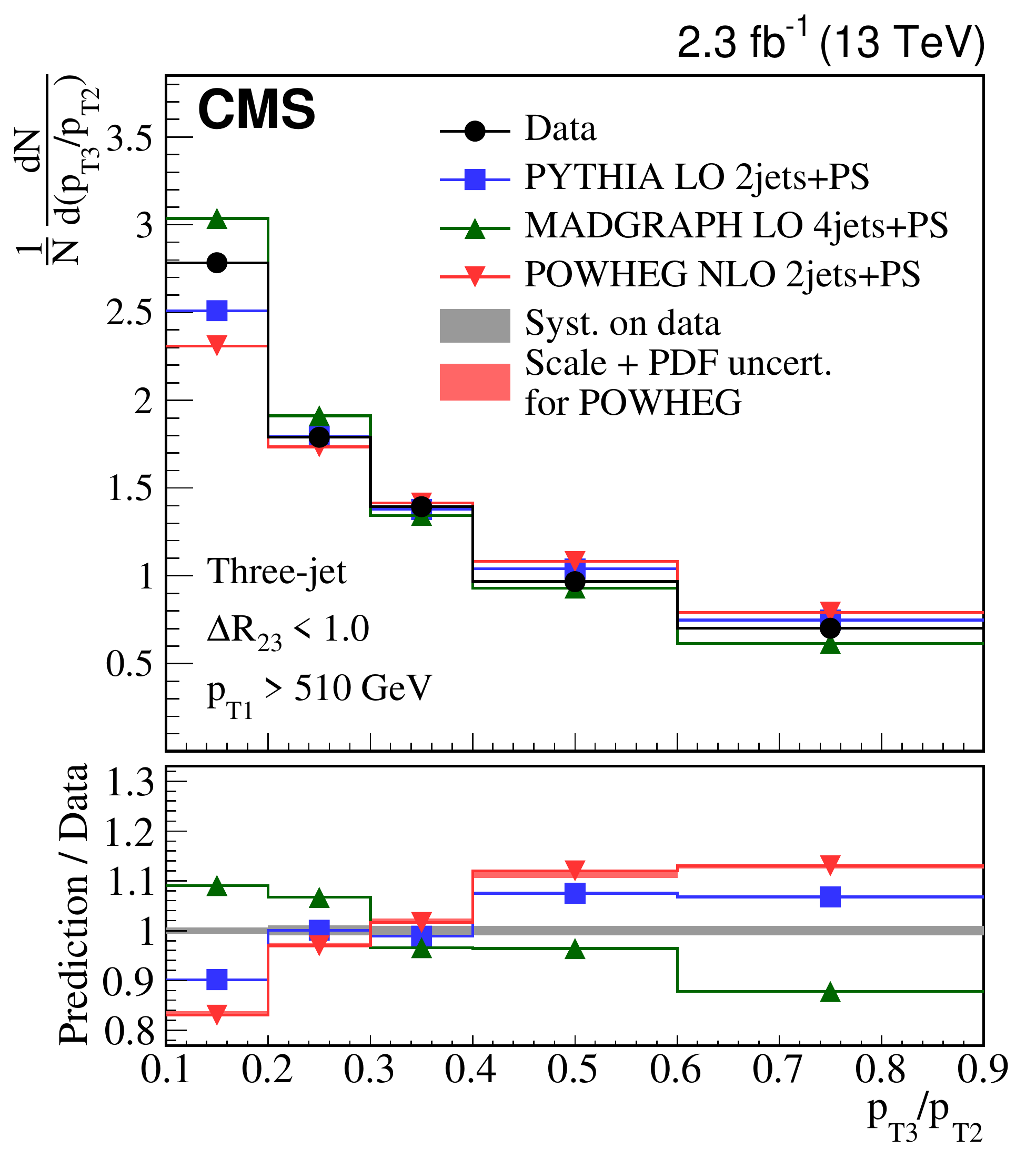} 
\includegraphics[width=0.4\textwidth]{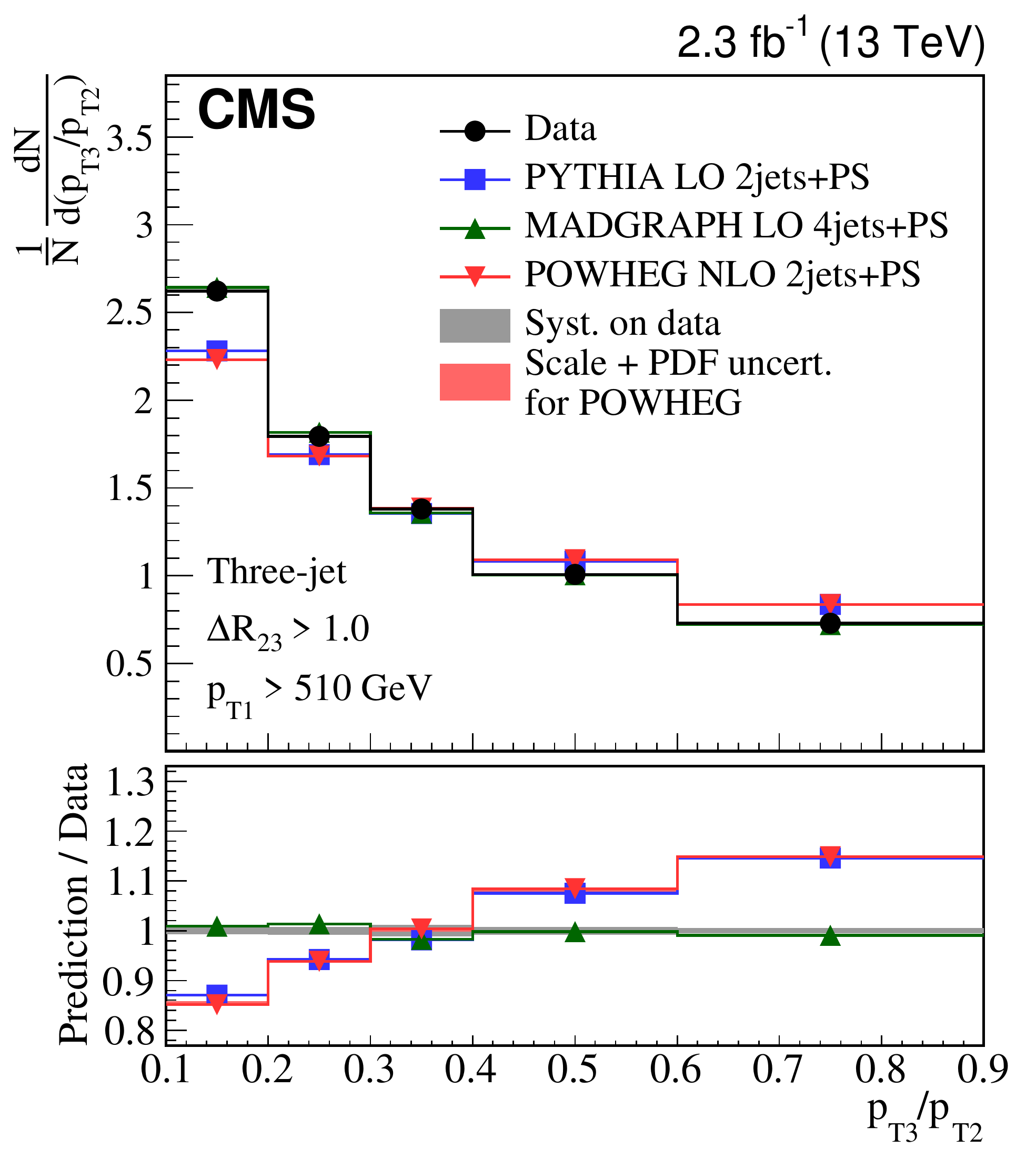}
\caption{\label{fig:13TeVpt} Three-jet events at $\sqrt{s} = 13\TeV$ compared to theory: (\cmsLeft) \protect\jetratio\ for small-angle radiation ($\deltaR < 1.0$), (\cmsRight) \protect\jetratio\ for large-angle radiation ($\deltaR > 1.0$).} 
\end{figure}

\begin{figure}[htb] 
\centering 
\includegraphics[width=0.4\textwidth]{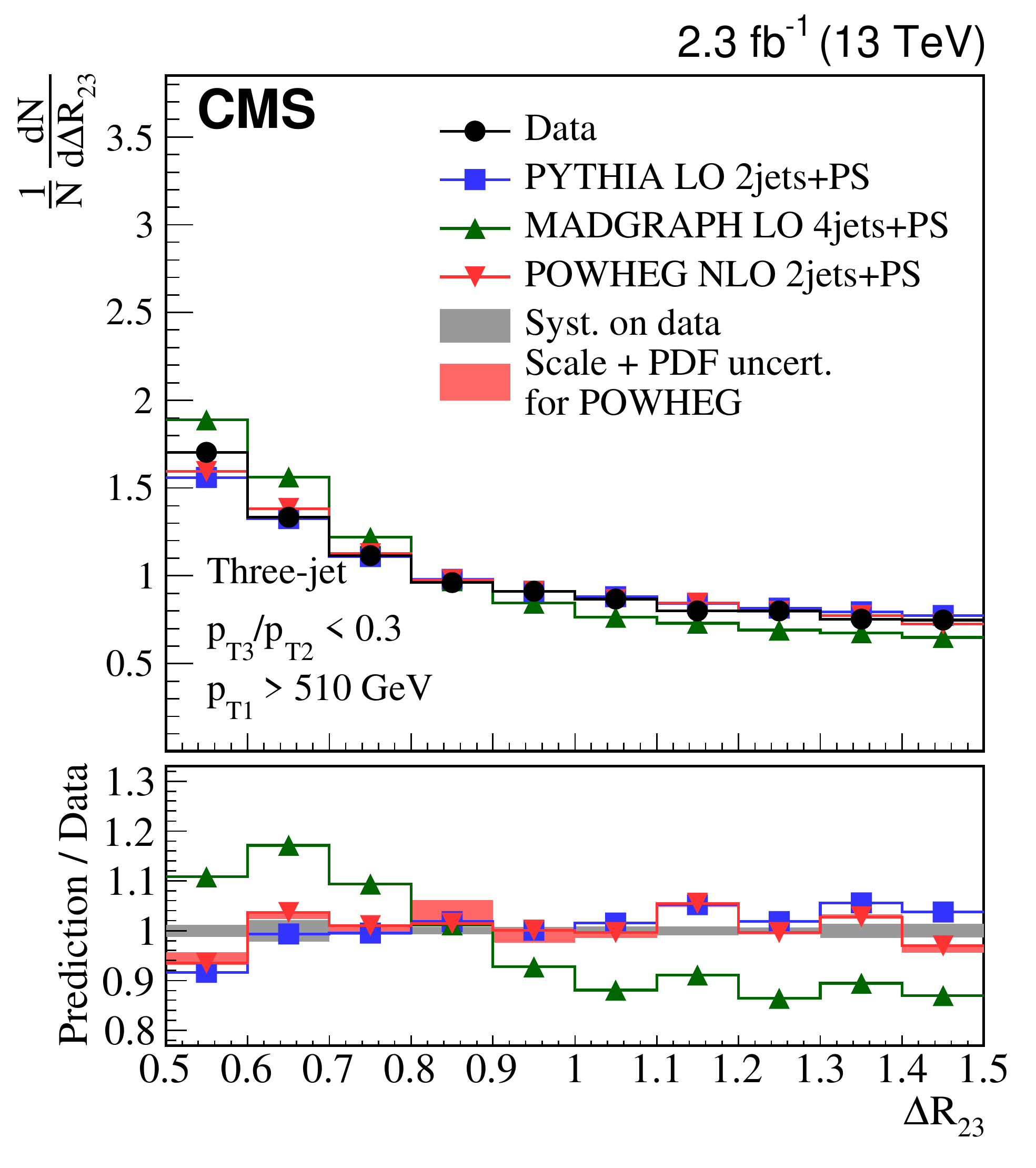} 
\includegraphics[width=0.4\textwidth]{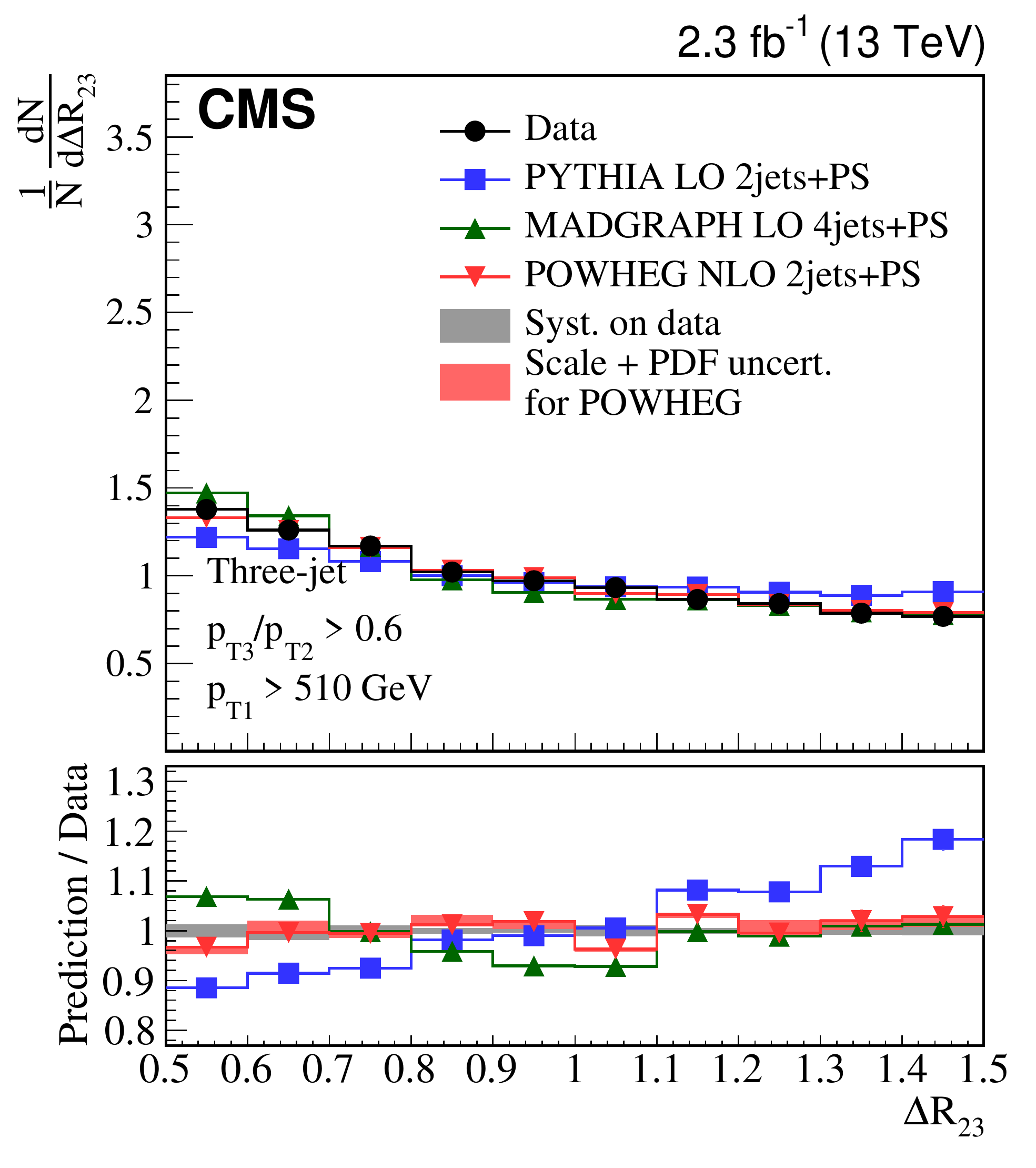}
\caption{\label{fig:13TeVDelta} Three-jet events at $\sqrt{s} = 13\TeV$ and comparison to theoretical predictions: (\cmsLeft) \deltaR\ for soft radiation ($\jetratio < 0.3$), (\cmsRight) \deltaR\ for hard radiation ($\jetratio > 0.6$).} 
\end{figure}

We show the $\sqrt{s}=8\TeV$ measurements of \jetratio\ in Fig.~\ref{fig:8TeVpt} and of \deltaR\ in Fig.~\ref{fig:8TeVDelta}, and compare them to theoretical expectations.
In Figs.~\ref{fig:13TeVpt} and \ref{fig:13TeVDelta} the distributions are given for $\sqrt{s} = 13\TeV$.
Figure~\ref{fig:8TeVpt} (\cmsLeft) shows the \jetratio\ distribution for the small \deltaR\ region.
All predictions show significant deviations from the measurements.
Interestingly, the \lofourps\ prediction shows different behavior compared with \lops\ and \nlops.
We see that the number of partons in the ME calculation and the merging method with the PS in the present simulations lead to different predictions. 
In Fig.~\ref{fig:8TeVpt} (\cmsRight) the \jetratio\ distribution is shown for large \deltaR.
This region of phase space is well described by the \lofourps~calculations, while the \lops\ and \nlops\ predictions show large deviations from the measurements.

In Fig.~\ref{fig:8TeVDelta}, the \deltaR\ distribution is shown for two regions of \jetratio .
Figure~\ref{fig:8TeVDelta} (\cmsLeft) shows $\jetratio < 0.3$.
The predictions from \lops\ and \nlops\ describe the measurement well, while the prediction from \lofourps\ shows a larger deviation from the data.
In Fig.~\ref{fig:8TeVDelta} (\cmsRight) the \deltaR\ distribution is shown for $\jetratio > 0.6$.
In contrast to Fig.~\ref{fig:8TeVDelta} (\cmsLeft), the predictions for distributions from \lops\ differ from the measurement, whereas the predictions from \nlops\ and \lofourps\ agree well with it.
This indicates that in this region the contribution from higher-multiplicity ME calculations supplemented with PS should be included.
The same comparisons are performed for the $\sqrt{s} = 13\TeV$ measurements as shown in Figs.~\ref{fig:13TeVpt} and \ref{fig:13TeVDelta}.
A similar behavior is observed for $\sqrt{s} = 8\TeV$.
In conclusion, none of the simulations simultaneously describes to simultaneously describe both the \jetratio\ and the \deltaR\ distributions in three-jet events.

\subsection{\texorpdfstring{$\PZ$}{Z} + two-jet selection}

\begin{figure}[htb] 
\centering
\includegraphics[width=0.4\textwidth]{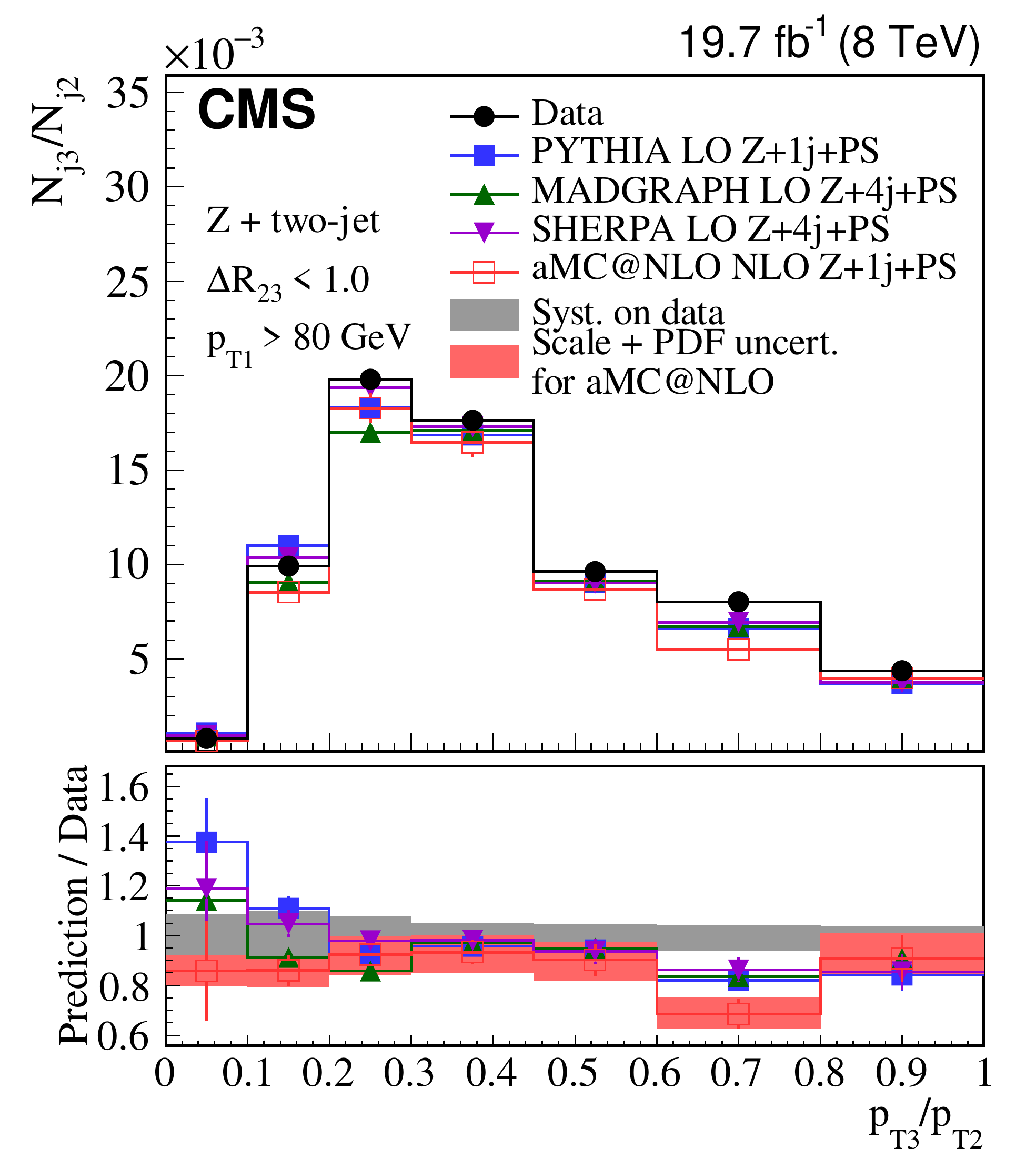} 
\includegraphics[width=0.4\textwidth]{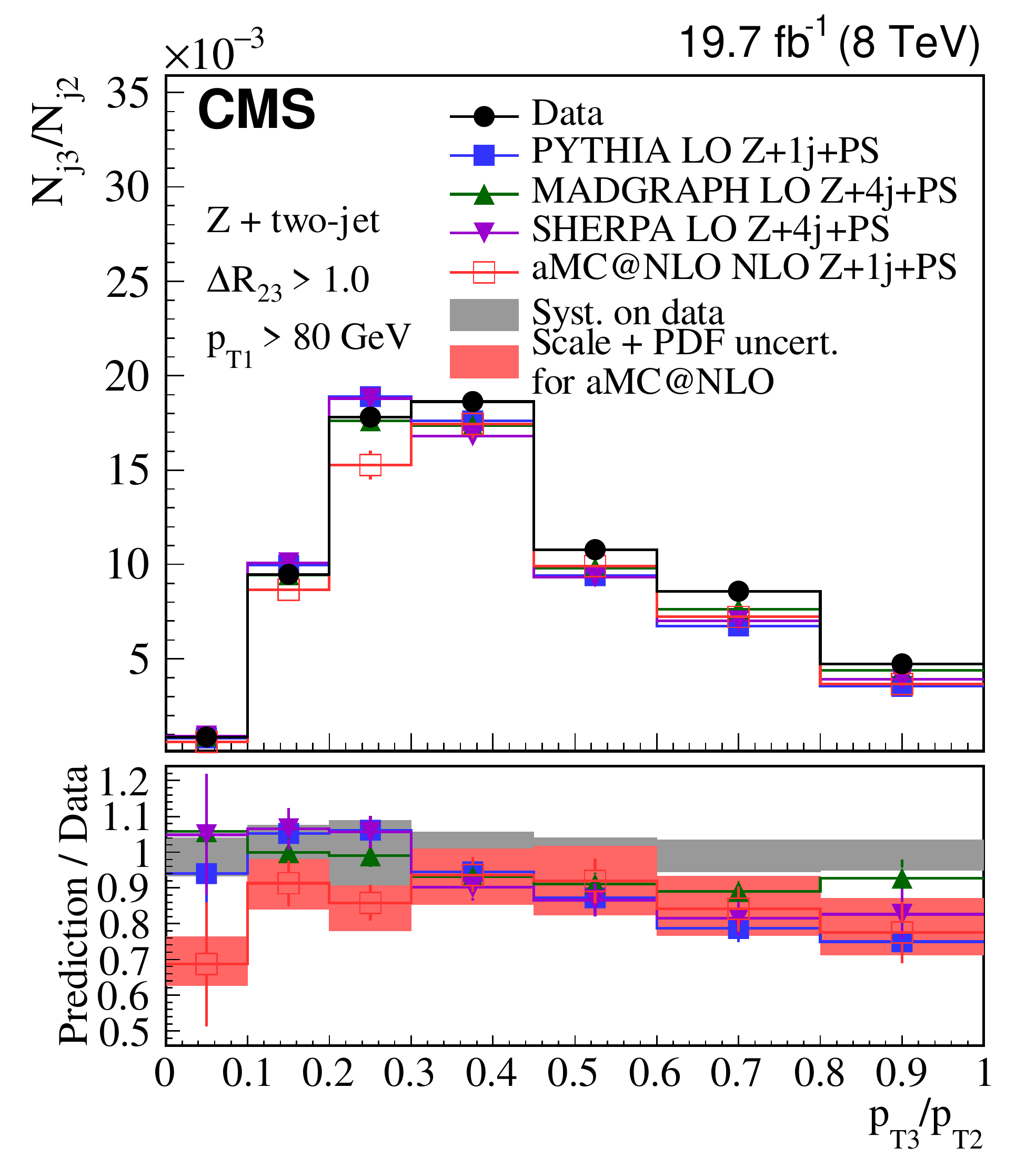}
\caption{\label{fig:Zfigpt} \Zjet\ events at $\sqrt{s} = 8\TeV$ compared to theory: (\cmsLeft) \protect\jetratio\ for small-angle radiation ($\deltaR < 1.0$), (\cmsRight) \protect\jetratio\ for large-angle radiation ($\deltaR > 1.0$).} 
\end{figure}

The measurement of \jetratio\ for \Zjet\ events is presented in Fig.~\ref{fig:Zfigpt} for data at $\sqrt{s} = 8\TeV$.
All distributions are normalized to the selected number of \PZ + one-jet events.
All predictions from \PYTHIA, \SHERPA, \MADGRAPH, and a\MCATNLO\ agree with data within the uncertainties of the measurement except for the phase space region with hard radiation.

\begin{figure}[htb] 
\centering 
\includegraphics[width=0.4\textwidth]{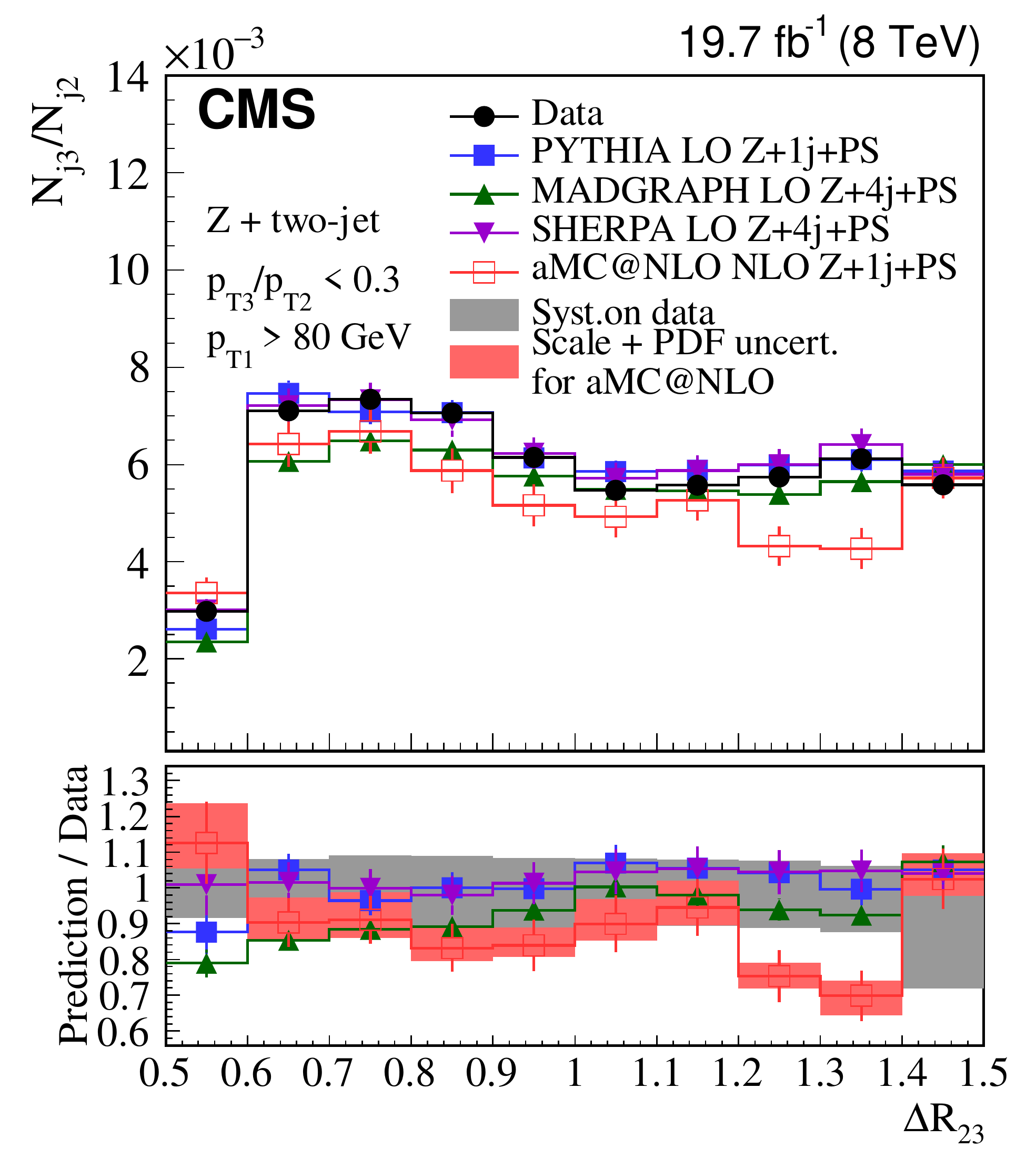} 
\includegraphics[width=0.4\textwidth]{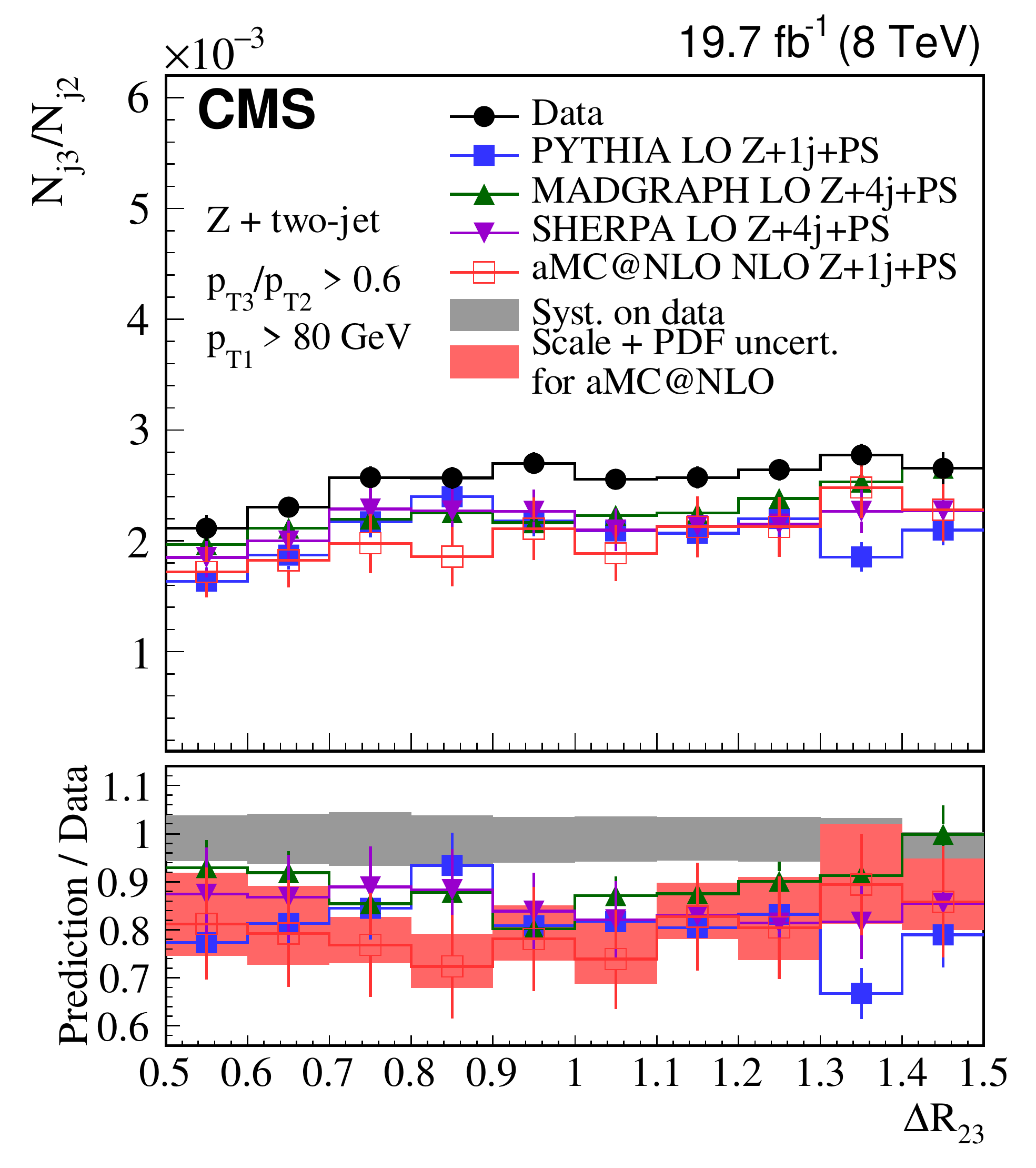}
\caption{\label{fig:ZfigDelta} \Zjet\ events at $\sqrt{s} = 8\TeV$ compared to theory: (\cmsLeft) \deltaR\ for soft radiation ($\jetratio < 0.3$), (\cmsRight) \deltaR\ for hard radiation ($\jetratio > 0.6$).} 
\end{figure}

Figure~\ref{fig:ZfigDelta} shows the measurement as a function of \deltaR.
The a\MCATNLO\ prediction deviates from the data at high \deltaR\ and small \jetratio, while \PYTHIA, \SHERPA, \MADGRAPH, and a\MCATNLO\ describe the shape of the distribution in the high-\jetratio\ range, but underestimate the data due to a smaller contribution from production of $j_3$.
This feature is based on the normalization of \Zjet\ distributions by the number of inclusive \PZ + one-jet events selected.

\begin{figure}[ht] 
\centering 
\includegraphics[width=0.4\textwidth]{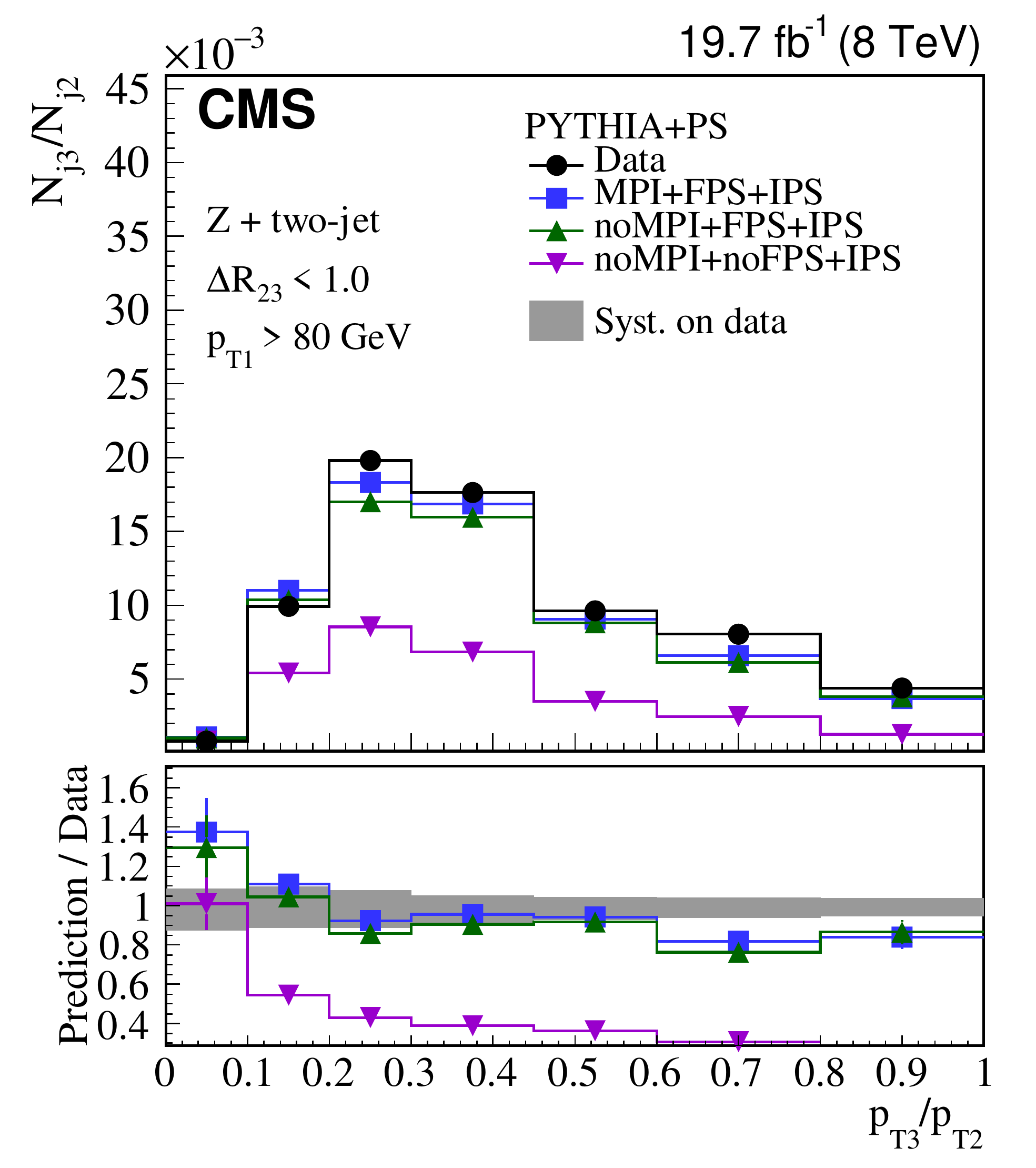} 
\includegraphics[width=0.4\textwidth]{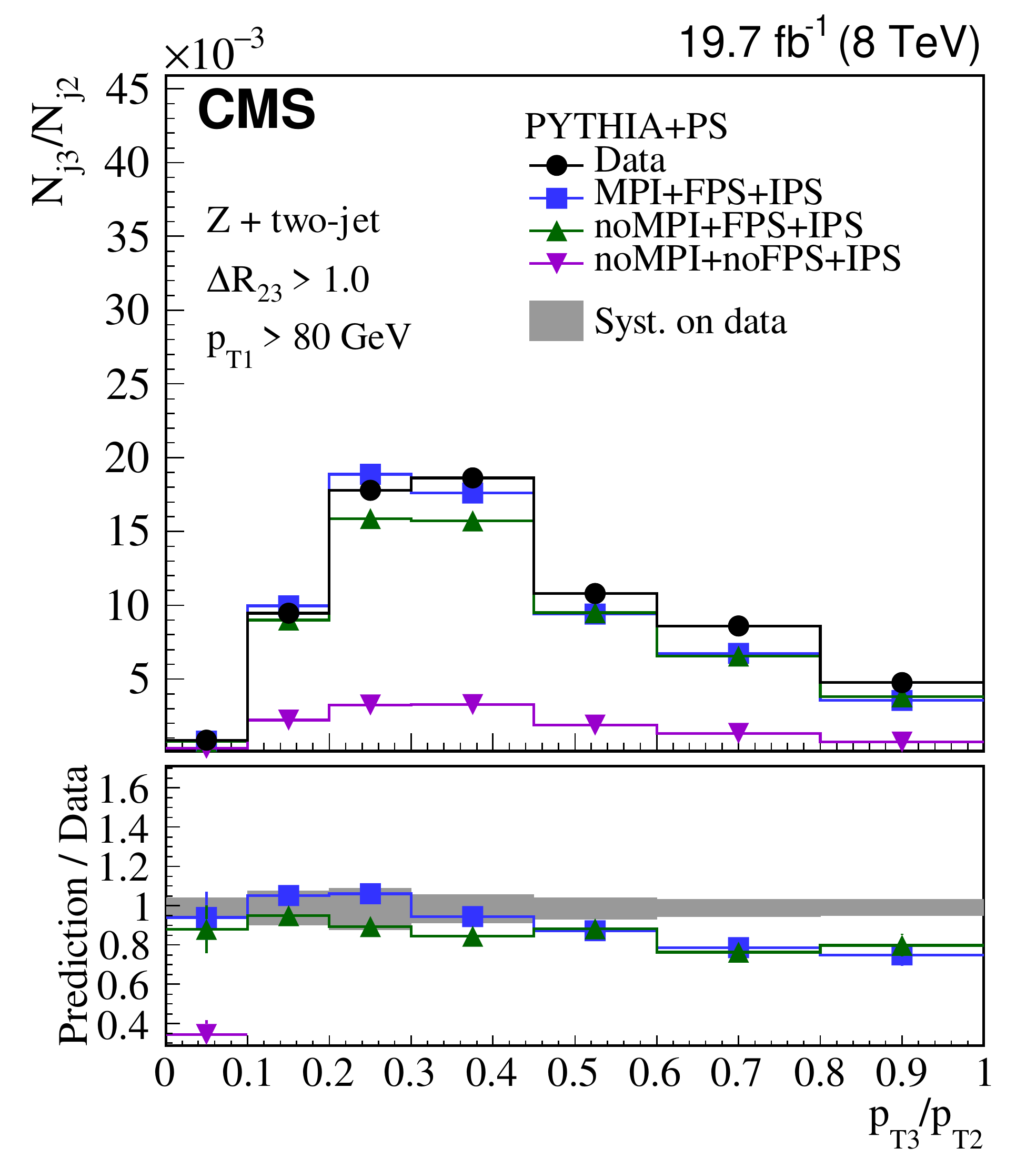}
\caption{\label{fig:Zfig6pt} \Zjet\ events at $\sqrt{s} = 8\TeV$ compared to theoretical predictions from \PYTHIA~8 without initial-state parton showers (IPS), final-state parton showers (FPS), and MPI: (\cmsLeft) \protect\jetratio\ for small-angle radiation ($\deltaR < 1.0$), (\cmsRight) \protect\jetratio\ for large-angle radiation ($\deltaR > 1.0$).} 
\end{figure}

\begin{figure}[ht] 
\centering 
\includegraphics[width=0.4\textwidth]{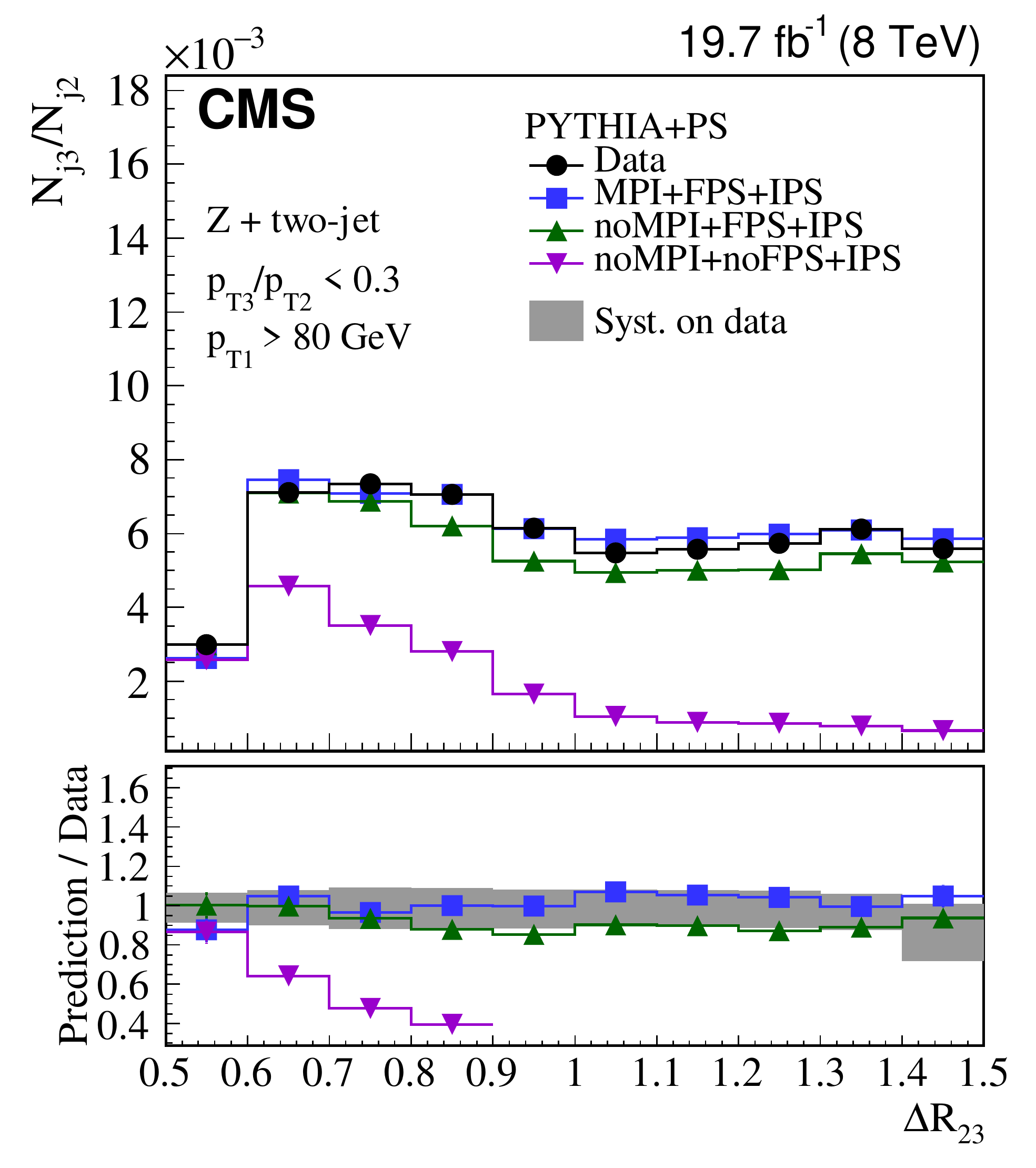} 
\includegraphics[width=0.4\textwidth]{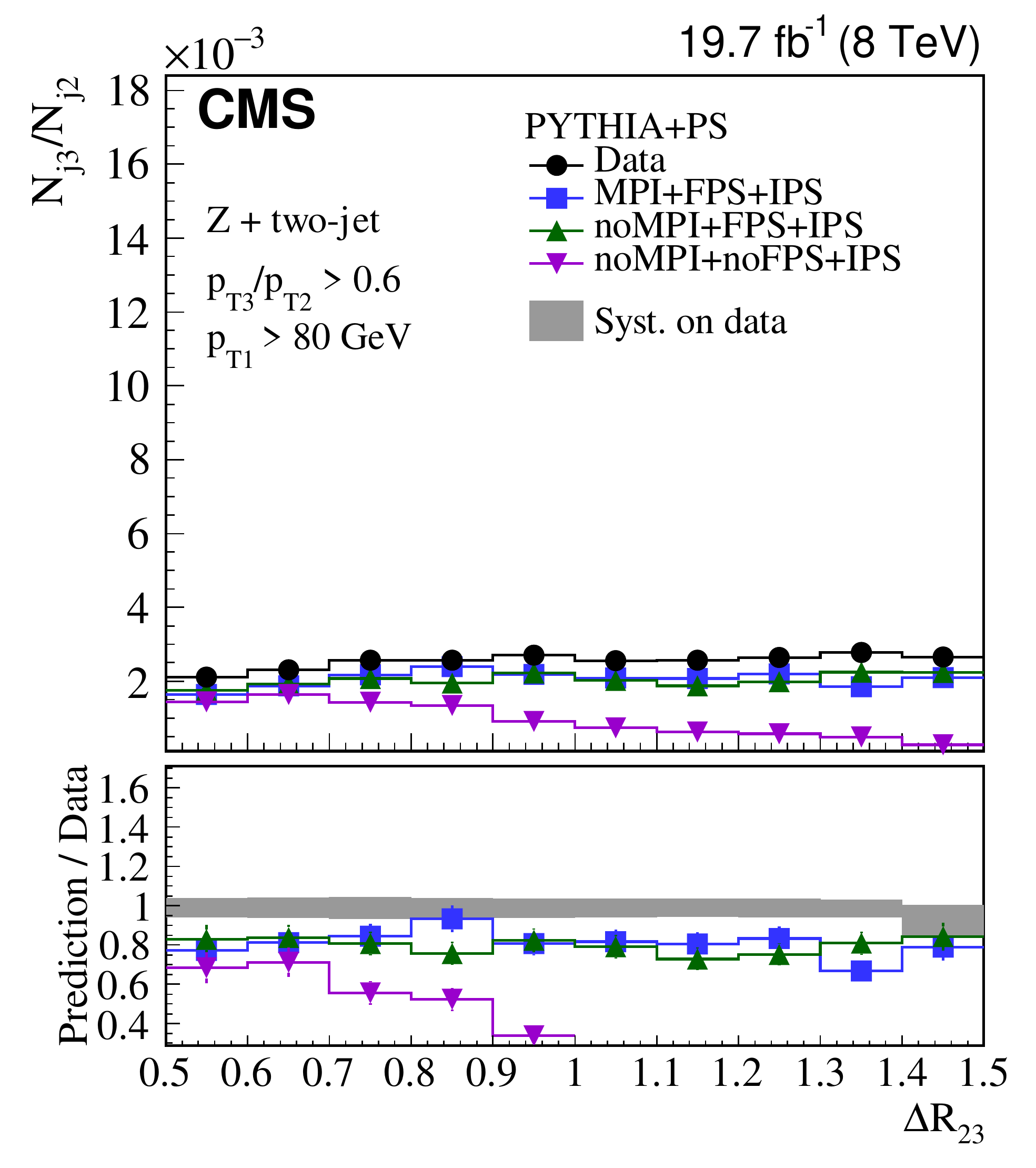}
\caption{\label{fig:Zfig6Delta} \Zjet\ events at $\sqrt{s} = 8\TeV$ and comparison to theoretical predictions from \PYTHIA~8 without initial-state parton showers (IPS), final-state parton showers (FPS), and MPI: (\cmsLeft) \deltaR\ for soft radiation ($\jetratio < 0.3$), (\cmsRight) \deltaR\ for hard radiation ($\jetratio > 0.6$).}
\end{figure}

Figures~\ref{fig:Zfig6pt} and \ref{fig:Zfig6Delta} compare the event distributions with predictions from \PYTHIA~8 with the final-state PS and MPI switched off.
The initial-state PS was kept, because one of the jets must originate from PS when \Zjet\ events are selected.
Multiple parton interactions play a very minor role, while the final-state PS in \PYTHIA~8 is very important.
When the final-state PS is switched off, events where both jets come from the initial-state PS are kept with a tendency to be close to each other in \deltaR.

In general, the measurements with \Zjet\ events are well described by all theoretical predictions, except for the underestimation of the $j_3$ emission.
The contribution of background from \ttbar production and dibosons can partially compensate the lack of the $j_3$ emission.
The contribution of the background (\ttbar production with fully leptonic decay and dibosons) increases the probability of $j_3$ emission from 2\% (soft radiation) to 10\% (hard radiation) depending on the phase space region.
The effect of the other processes (\ttbar production with semileptonic and hadronic decays, single top production) is negligible.
In comparison with the three-jet measurements, we observe significant differences; only in the region of large \deltaR\ and large \jetratio\ (hard and large-angle radiation) do the theoretical predictions agree with the measurement.
The accessible range in \pt is rather small in \Zjet\ events because of the limit in the \pt of the \PZ bosons ($\ptl > 80\GeV$), while the three-jet selection, on the contrary, can have a rather large range ($\ptl > 510\GeV$).
This may explain why the region of small \jetratio\ is better described by predictions that include PS in the latter case.
In addition, the large-angle radiation is best described by fixed-order ME calculations.

In conclusion, the \Zjet\ measurement has a different distribution in \jetratio, which originates from the different kinematic selection criteria relative to three-jet events, thus reducing the sensitivity in the soft and collinear region.
Within the available phase space, the measurements are in reasonable agreement with both PS and ME calculations, apart from the emission of $j_3$ in the high-\jetratio\ region.

\section{Summary}

Two kinematic variables are introduced to quantify the radiation pattern in multijet events: (i) the transverse momentum ratio (\jetratio) of two jets, and (ii) their angular separation (\deltaR). The variable \jetratio\ is used to distinguish between soft and hard radiation, while \deltaR\ classifies events into small- and large-angle radiation types.
Events with three or more energetic jets as well as inclusive \Zjet\ events are selected for study using data collected at $\sqrt{s} = 8\TeV$ corresponding to an integrated luminosity of 19.8\fbinv.
Three-jet events at $\sqrt{s} = 13\TeV$ corresponding to an integrated luminosity of 2.3\fbinv are also analyzed.
No significant dependence on the center-of-mass energy is observed in the differential distributions of \jetratio\ and \deltaR.

{\tolerance=800
Overall, large-angle radiation (large \deltaR) and hard radiation (large \jetratio) are well described by the matrix element (ME) calculations (using \lofourps\ formulations), while the parton shower (PS) approach (\lops\ and \nlops) fail to describe the regions of large-angle and hard radiation. 
The collinear region (small \deltaR) is not well described; \lops, \nlops, and \lofourps\ distributions show deviations from the measurements.
In the soft region (small \jetratio), the PS approach describes the measurement also in the large-angle region (full range in \deltaR), while for large \jetratio\ higher-order ME contributions are needed to describe the three-jet measurements.
The distributions in \Zjet\ events are reasonably described by all tested generators.
Nevertheless, we find an underestimation of third-jet emission at large \jetratio\ both in the collinear and large-angle regions, for all of the tested models.
Contribution from \ttbar and dibosons production may partially cover the difference.
These results illustrate how well the collinear/soft, and large-angle/hard regions are described by different approaches.
The different kinematic regions and initial-state flavor composition may be the reason why the three-jet measurements are less consistent with the theoretical predictions relative to the \Zjet\ final states. 
These results clearly indicate that the methods of merging ME with PS calculations are not yet optimal for describing the full region of phase space.
The measurements presented here serve as benchmarks for future improved predictions coming from ME calcualtions combined with parton showers.
\par}

\begin{acknowledgments}
    We congratulate our colleagues in the CERN accelerator departments for the excellent performance of the LHC and thank the technical and administrative staffs at CERN and at other CMS institutes for their contributions to the success of the CMS effort. In addition, we gratefully acknowledge the computing centers and personnel of the Worldwide LHC Computing Grid and other centers for delivering so effectively the computing infrastructure essential to our analyses. Finally, we acknowledge the enduring support for the construction and operation of the LHC, the CMS detector, and the supporting computing infrastructure provided by the following funding agencies: BMBWF and FWF (Austria); FNRS and FWO (Belgium); CNPq, CAPES, FAPERJ, FAPERGS, and FAPESP (Brazil); MES (Bulgaria); CERN; CAS, MoST, and NSFC (China); COLCIENCIAS (Colombia); MSES and CSF (Croatia); RIF (Cyprus); SENESCYT (Ecuador); MoER, ERC PUT and ERDF (Estonia); Academy of Finland, MEC, and HIP (Finland); CEA and CNRS/IN2P3 (France); BMBF, DFG, and HGF (Germany); GSRT (Greece); NKFIA (Hungary); DAE and DST (India); IPM (Iran); SFI (Ireland); INFN (Italy); MSIP and NRF (Republic of Korea); MES (Latvia); LAS (Lithuania); MOE and UM (Malaysia); BUAP, CINVESTAV, CONACYT, LNS, SEP, and UASLP-FAI (Mexico); MOS (Montenegro); MBIE (New Zealand); PAEC (Pakistan); MSHE and NSC (Poland); FCT (Portugal); JINR (Dubna); MON, RosAtom, RAS, RFBR, and NRC KI (Russia); MESTD (Serbia); SEIDI, CPAN, PCTI, and FEDER (Spain); MOSTR (Sri Lanka); Swiss Funding Agencies (Switzerland); MST (Taipei); ThEPCenter, IPST, STAR, and NSTDA (Thailand); TUBITAK and TAEK (Turkey); NASU (Ukraine); STFC (United Kingdom); DOE and NSF (USA).

    \hyphenation{Rachada-pisek} Individuals have received support from the Marie-Curie program and the European Research Council and Horizon 2020 Grant, contract Nos.\ 675440, 724704, 752730, and 765710 (European Union); the Leventis Foundation; the Alfred P.\ Sloan Foundation; the Alexander von Humboldt Foundation; the Belgian Federal Science Policy Office; the Fonds pour la Formation \`a la Recherche dans l'Industrie et dans l'Agriculture (FRIA-Belgium); the Agentschap voor Innovatie door Wetenschap en Technologie (IWT-Belgium); the F.R.S.-FNRS and FWO (Belgium) under the ``Excellence of Science -- EOS" -- be.h project n.\ 30820817; the Beijing Municipal Science \& Technology Commission, No. Z191100007219010; the Ministry of Education, Youth and Sports (MEYS) of the Czech Republic; the Deutsche Forschungsgemeinschaft (DFG), under Germany's Excellence Strategy -- EXC 2121 ``Quantum Universe" -- 390833306, and under project number 400140256 - GRK2497; the Lend\"ulet (``Momentum") Program and the J\'anos Bolyai Research Scholarship of the Hungarian Academy of Sciences, the New National Excellence Program \'UNKP, the NKFIA research grants 123842, 123959, 124845, 124850, 125105, 128713, 128786, and 129058 (Hungary); the Council of Science and Industrial Research, India; the HOMING PLUS program of the Foundation for Polish Science, cofinanced from European Union, Regional Development Fund, the Mobility Plus program of the Ministry of Science and Higher Education, the National Science Center (Poland), contracts Harmonia 2014/14/M/ST2/00428, Opus 2014/13/B/ST2/02543, 2014/15/B/ST2/03998, and 2015/19/B/ST2/02861, Sonata-bis 2012/07/E/ST2/01406; the National Priorities Research Program by Qatar National Research Fund; the Ministry of Science and Higher Education, project no. 0723-2020-0041 (Russia); the Programa Estatal de Fomento de la Investigaci{\'o}n Cient{\'i}fica y T{\'e}cnica de Excelencia Mar\'{\i}a de Maeztu, grant MDM-2015-0509 and the Programa Severo Ochoa del Principado de Asturias; the Thalis and Aristeia programs cofinanced by EU-ESF and the Greek NSRF; the Rachadapisek Sompot Fund for Postdoctoral Fellowship, Chulalongkorn University and the Chulalongkorn Academic into Its 2nd Century Project Advancement Project (Thailand); the Kavli Foundation; the Nvidia Corporation; the SuperMicro Corporation; the Welch Foundation, contract C-1845; and the Weston Havens Foundation (USA).
\end{acknowledgments}

\bibliography{auto_generated}  

\providecommand{\href}[2]{#2}\begingroup\raggedright\begin{thebibliography}{10}%
\makeatletter
\providecommand{\hrefCMSnoop }[0]{\@secondoftwo}%
\makeatother
\providecommand{\doi}{\texttt{doi:}\begingroup \urlstyle{tt}\Url}

\bibitem{Catani:2001cc}
\hrefCMSnoop {}{S.~Catani, F.~Krauss, R.~Kuhn, and B.~R. Webber, ``{QCD matrix
  elements + parton showers}'',} \textit{ JHEP} \textbf{ 11} (2001) 063,
  \href{http://dx.doi.org/10.1088/1126-6708/2001/11/063}{\doi{10.1088/1126-6708/2001/11/063}},
  \href{http://www.arXiv.org/abs/hep-ph/0109231}{\texttt{arXiv:hep-ph/0109231}}.

\bibitem{Buckley:2011ms}
A.~Buckley\hrefCMSnoop {}{ {et~al.}, ``{General-purpose event generators for
  LHC physics}'',} \textit{ Phys. Rept.} \textbf{ 504} (2011) 145,
  \href{http://dx.doi.org/10.1016/j.physrep.2011.03.005}{\doi{10.1016/j.physrep.2011.03.005}},
\href{http://www.arXiv.org/abs/1101.2599}{\texttt{arXiv:1101.2599}}.

\bibitem{Bengtsson:1986hr}
\hrefCMSnoop {}{M.~Bengtsson and T.~Sj{\"o}strand, ``{Coherent parton showers
  versus matrix elements-implications of PETRA/PEP data}'',} \textit{ Phys.
  Lett. B} \textbf{ 185} (1987) 435,
  \href{http://dx.doi.org/10.1016/0370-2693(87)91031-8}{\doi{10.1016/0370-2693(87)91031-8}}.

\bibitem{Mrenna:2003if}
\hrefCMSnoop {}{S.~Mrenna and P.~Richardson, ``{Matching matrix elements and
  parton showers with HERWIG and PYTHIA}'',} \textit{ JHEP} \textbf{ 05} (2004)
  040,
  \href{http://dx.doi.org/10.1088/1126-6708/2004/05/040}{\doi{10.1088/1126-6708/2004/05/040}},
\href{http://www.arXiv.org/abs/hep-ph/0312274}{\texttt{arXiv:hep-ph/0312274}}.

\bibitem{Chatrchyan:2013fha}
\hrefCMSnoop {}{{CMS Collaboration}, ``{Probing color coherence effects in pp
  collisions at $\sqrt{s}=7\,\text {TeV} $}'',} \textit{ Eur. Phys. J. C}
  \textbf{ 74} (2014) 2901,
  \href{http://dx.doi.org/10.1140/epjc/s10052-014-2901-8}{\doi{10.1140/epjc/s10052-014-2901-8}},
\href{http://www.arXiv.org/abs/1311.5815}{\texttt{arXiv:1311.5815}}.

\bibitem{Abe:1994nj}
\hrefCMSnoop {}{{CDF} Collaboration, ``{Evidence for color coherence in
  $\rm{p\bar{p}}$ collisions at $\sqrt{s} = 1.8$ TeV}'',} \textit{ Phys. Rev.
  D} \textbf{ 50} (1994) 5562,
\href{http://dx.doi.org/10.1103/PhysRevD.50.5562}{\doi{10.1103/PhysRevD.50.5562}}.

\bibitem{Abbott:1997bk}
\hrefCMSnoop {}{{D0} Collaboration, ``{Color coherent radiation in multijet
  events from $\rm{p\bar{p}}$ collisions at $\sqrt{s} = 1.8$ TeV}'',} \textit{
  Phys. Lett. B} \textbf{ 414} (1997) 419,
  \href{http://dx.doi.org/10.1016/S0370-2693(97)01190-8}{\doi{10.1016/S0370-2693(97)01190-8}},
\href{http://www.arXiv.org/abs/hep-ex/9706012}{\texttt{arXiv:hep-ex/9706012}}.

\bibitem{Chatrchyan:2008zzk}
\hrefCMSnoop {}{{CMS Collaboration}, ``The {CMS} experiment at the {CERN}
  {LHC}'',} \textit{ JINST} \textbf{ 03} (2008) S08004,
  \href{http://dx.doi.org/10.1088/1748-0221/3/08/S08004}{\doi{10.1088/1748-0221/3/08/S08004}}.

\bibitem{Khachatryan:2016bia}
\hrefCMSnoop {}{{CMS Collaboration}, ``{The CMS trigger system}'',} \textit{
  JINST} \textbf{ 12} (2017) P01020,
  \href{http://dx.doi.org/10.1088/1748-0221/12/01/P01020}{\doi{10.1088/1748-0221/12/01/P01020}},
  \href{http://www.arXiv.org/abs/1609.02366}{\texttt{arXiv:1609.02366}}.

\bibitem{bib:parflow}
\hrefCMSnoop {}{{CMS Collaboration}, ``{Particle-flow reconstruction and global
  event description with the CMS detector}'',} \textit{ JINST} \textbf{ 12}
  (2017) P10003,
  \href{http://dx.doi.org/10.1088/1748-0221/12/10/P10003}{\doi{10.1088/1748-0221/12/10/P10003}},
  \href{http://www.arXiv.org/abs/1706.04965}{\texttt{arXiv:1706.04965}}.

\bibitem{Cacciari:2008gp}
\hrefCMSnoop {}{M.~Cacciari, G.~P. Salam, and G.~Soyez, ``The anti-\kt jet
  clustering algorithm'',} \textit{ JHEP} \textbf{ 04} (2008) 063,
  \href{http://dx.doi.org/10.1088/1126-6708/2008/04/063}{\doi{10.1088/1126-6708/2008/04/063}},
\href{http://www.arXiv.org/abs/0802.1189}{\texttt{arXiv:0802.1189}}.

\bibitem{Cacciari:2011ma}
\hrefCMSnoop {}{M.~Cacciari, G.~P. Salam, and G.~Soyez, ``{FastJet} user
  manual'',} \textit{ Eur. Phys. J. C} \textbf{ 72} (2012) 1896,
  \href{http://dx.doi.org/10.1140/epjc/s10052-012-1896-2}{\doi{10.1140/epjc/s10052-012-1896-2}},
  \href{http://www.arXiv.org/abs/1111.6097}{\texttt{arXiv:1111.6097}}.

\bibitem{Chatrchyan:2013sba}
\hrefCMSnoop {}{{CMS Collaboration}, ``{The performance of the CMS muon
  detector in proton-proton collisions at $\sqrt{s} = 7$ TeV at the LHC}'',}
  \textit{ JINST} \textbf{ 8} (2013) P11002,
  \href{http://dx.doi.org/10.1088/1748-0221/8/11/P11002}{\doi{10.1088/1748-0221/8/11/P11002}},
\href{http://www.arXiv.org/abs/1306.6905}{\texttt{arXiv:1306.6905}}.

\bibitem{Sirunyan:2018fpa}
\hrefCMSnoop {}{{CMS Collaboration}, ``{Performance of the CMS muon detector
  and muon reconstruction with proton-proton collisions at $\sqrt{s}=$ 13
  TeV}'',} \textit{ JINST} \textbf{ 13} (2018) P06015,
  \href{http://dx.doi.org/10.1088/1748-0221/13/06/P06015}{\doi{10.1088/1748-0221/13/06/P06015}},
\href{http://www.arXiv.org/abs/1804.04528}{\texttt{arXiv:1804.04528}}.

\bibitem{Chatrchyan:2012xi}
\hrefCMSnoop {}{{CMS Collaboration}, ``{Performance of CMS muon reconstruction
  in pp collision events at $\sqrt{s}=7$ TeV}'',} \textit{ JINST} \textbf{ 7}
  (2012) P10002,
  \href{http://dx.doi.org/10.1088/1748-0221/7/10/P10002}{\doi{10.1088/1748-0221/7/10/P10002}},
\href{http://www.arXiv.org/abs/1206.4071}{\texttt{arXiv:1206.4071}}.

\bibitem{CMS-DP-2013-009}
\href {https://cds.cern.ch/record/1536406}{{CMS Collaboration}, ``{Single Muon
  efficiencies in 2012 Data}'',} CMS Detector Performance Note CMS-DP-2013-009,
  2013.

\bibitem{Khachatryan:2016mlc}
\hrefCMSnoop {}{{CMS Collaboration}, ``{Measurement and QCD analysis of
  double-differential inclusive jet cross sections in pp collisions at $
  \sqrt{s}=8 $ TeV and cross section ratios to 2.76 and 7 TeV}'',} \textit{
  JHEP} \textbf{ 03} (2017) 156,
  \href{http://dx.doi.org/10.1007/JHEP03(2017)156}{\doi{10.1007/JHEP03(2017)156}},
\href{http://www.arXiv.org/abs/1609.05331}{\texttt{arXiv:1609.05331}}.

\bibitem{Khachatryan:2016wdh}
\hrefCMSnoop {}{{CMS Collaboration}, ``{Measurement of the double-differential
  inclusive jet cross section in proton-proton collisions at $\sqrt{s} =
  13\,\text {TeV} $}'',} \textit{ Eur. Phys. J. C} \textbf{ 76} (2016) 451,
  \href{http://dx.doi.org/10.1140/epjc/s10052-016-4286-3}{\doi{10.1140/epjc/s10052-016-4286-3}},
  \href{http://www.arXiv.org/abs/1605.04436}{\texttt{arXiv:1605.04436}}.

\bibitem{CMS-PAPERS-JME-10-009}
\hrefCMSnoop {}{{CMS Collaboration}, ``{Missing transverse energy performance
  of the {CMS} detector}'',} \textit{ JINST} \textbf{ 6} (2011) P09001,
  \href{http://dx.doi.org/10.1088/1748-0221/6/09/P09001}{\doi{10.1088/1748-0221/6/09/P09001}},
  \href{http://www.arXiv.org/abs/1106.5048}{\texttt{arXiv:1106.5048}}.

\bibitem{CMS-PAPERS-JME-13-003}
\hrefCMSnoop {}{{CMS Collaboration}, ``{Performance of the {CMS} missing
  transverse momentum reconstruction in pp data at {$\sqrt{s}$} = 8 {TeV}}'',}
  \textit{ JINST} \textbf{ 10} (2015) P02006,
  \href{http://dx.doi.org/10.1088/1748-0221/10/02/P02006}{\doi{10.1088/1748-0221/10/02/P02006}},
  \href{http://www.arXiv.org/abs/1411.0511}{\texttt{arXiv:1411.0511}}.

\bibitem{CMS-PAPERS-JME-17-001}
\hrefCMSnoop {}{{CMS Collaboration}, ``{Performance of missing transverse
  momentum reconstruction in proton-proton collisions at {$\sqrt{s} =$} 13
  {TeV} using the {CMS} detector}'',} \textit{ JINST} \textbf{ 14} (2019)
  P07004,
  \href{http://dx.doi.org/10.1088/1748-0221/14/07/P07004}{\doi{10.1088/1748-0221/14/07/P07004}},
  \href{http://www.arXiv.org/abs/1903.06078}{\texttt{arXiv:1903.06078}}.

\bibitem{bib:geant}
\hrefCMSnoop {}{{GEANT4} Collaboration, ``{\GEANTfour}---a simulation
  toolkit'',} \textit{ Nucl. Instrum. Meth. A} \textbf{ 506} (2003) 250,
\href{http://dx.doi.org/10.1016/S0168-9002(03)01368-8}{\doi{10.1016/S0168-9002(03)01368-8}}.

\bibitem{bib:madgraph5}
J.~Alwall\hrefCMSnoop {}{ {et~al.}, ``The automated computation of tree-level
  and next-to-leading order differential cross sections, and their matching to
  parton shower simulations'',} \textit{ JHEP} \textbf{ 07} (2014) 079,
  \href{http://dx.doi.org/10.1007/JHEP07(2014)079}{\doi{10.1007/JHEP07(2014)079}},
  \href{http://www.arXiv.org/abs/1405.0301}{\texttt{arXiv:1405.0301}}.

\bibitem{Sjostrand:2006za}
\hrefCMSnoop {}{T.~Sj{\"o}strand, S.~Mrenna, and P.~Skands, ``{PYTHIA} 6.4
  physics and manual'',} \textit{ JHEP} \textbf{ 05} (2006) 026,
  \href{http://dx.doi.org/10.1088/1126-6708/2006/05/026}{\doi{10.1088/1126-6708/2006/05/026}},
\href{http://www.arXiv.org/abs/hep-ph/0603175}{\texttt{arXiv:hep-ph/0603175}}.

\bibitem{Pumplin:2002vw}
J.~Pumplin\hrefCMSnoop {}{ {et~al.}, ``New generation of parton distributions
  with uncertainties from global {QCD} analysis'',} \textit{ JHEP} \textbf{ 07}
  (2002) 012,
  \href{http://dx.doi.org/10.1088/1126-6708/2002/07/012}{\doi{10.1088/1126-6708/2002/07/012}},
\href{http://www.arXiv.org/abs/hep-ph/0201195}{\texttt{arXiv:hep-ph/0201195}}.

\bibitem{CMS-PAPERS-QCD-10-010}
\hrefCMSnoop {}{{CMS Collaboration}, ``Measurement of the underlying event
  activity at the {LHC} with {$\sqrt{s} = 7\TeV$} and comparison with
  {$\sqrt{s} = 0.9\TeV$}'',} \textit{ JHEP} \textbf{ 09} (2011) 109,
  \href{http://dx.doi.org/10.1007/JHEP09(2011)109}{\doi{10.1007/JHEP09(2011)109}},
  \href{http://www.arXiv.org/abs/1107.0330}{\texttt{arXiv:1107.0330}}.

\bibitem{Sjostrand:2007gs}
\hrefCMSnoop {}{T.~Sj{\"o}strand, S.~Mrenna, and P.~Skands, ``A brief
  introduction to {PYTHIA} 8.1'',} \textit{ Comp. Phys. Comm.} \textbf{ 178}
  (2008) 852,
  \href{http://dx.doi.org/10.1016/j.cpc.2008.01.036}{\doi{10.1016/j.cpc.2008.01.036}},
\href{http://www.arXiv.org/abs/0710.3820}{\texttt{arXiv:0710.3820}}.

\bibitem{Corke_2011}
\hrefCMSnoop {}{R.~Corke and T.~Sj{\"o}strand, ``{Interleaved Parton Showers
  and Tuning Prospects}'',} \textit{ JHEP} \textbf{ 03} (2011) 032,
  \href{http://dx.doi.org/10.1007/JHEP03(2011)032}{\doi{10.1007/JHEP03(2011)032}},
  \href{http://www.arXiv.org/abs/1011.1759}{\texttt{arXiv:1011.1759}}.

\bibitem{Sjostrand:2014zea}
T.~Sj{\"o}strand\hrefCMSnoop {}{ {et~al.}, ``An introduction to {PYTHIA}
  8.2'',} \textit{ Comput. Phys. Commun.} \textbf{ 191} (2015) 159,
  \href{http://dx.doi.org/10.1016/j.cpc.2015.01.024}{\doi{10.1016/j.cpc.2015.01.024}},
  \href{http://www.arXiv.org/abs/1410.3012}{\texttt{arXiv:1410.3012}}.

\bibitem{Ball:2012cx}
\hrefCMSnoop {}{{NNPDF} Collaboration, ``{Parton distributions with LHC
  data}'',} \textit{ Nucl. Phys. B} \textbf{ 867} (2013) 244,
  \href{http://dx.doi.org/10.1016/j.nuclphysb.2012.10.003}{\doi{10.1016/j.nuclphysb.2012.10.003}},
\href{http://www.arXiv.org/abs/1207.1303}{\texttt{arXiv:1207.1303}}.

\bibitem{Khachatryan:2015pea}
\hrefCMSnoop {}{{CMS Collaboration}, ``{Event generator tunes obtained from
  underlying event and multiparton scattering measurements}'',} \textit{ Eur.
  Phys. J. C} \textbf{ 76} (2016) 155,
  \href{http://dx.doi.org/10.1140/epjc/s10052-016-3988-x}{\doi{10.1140/epjc/s10052-016-3988-x}},
\href{http://www.arXiv.org/abs/1512.00815}{\texttt{arXiv:1512.00815}}.

\bibitem{Gleisberg:2008ta}
T.~Gleisberg\hrefCMSnoop {}{ {et~al.}, ``Event generation with {SHERPA} 1.1'',}
  \textit{ JHEP} \textbf{ 02} (2009) 007,
  \href{http://dx.doi.org/10.1088/1126-6708/2009/02/007}{\doi{10.1088/1126-6708/2009/02/007}},
  \href{http://www.arXiv.org/abs/0811.4622}{\texttt{arXiv:0811.4622}}.

\bibitem{Schumann:2007mg}
\hrefCMSnoop {}{S.~Schumann and F.~Krauss, ``{A parton shower algorithm based
  on Catani-Seymour dipole factorisation}'',} \textit{ JHEP} \textbf{ 03}
  (2008) 038,
  \href{http://dx.doi.org/10.1088/1126-6708/2008/03/038}{\doi{10.1088/1126-6708/2008/03/038}},
  \href{http://www.arXiv.org/abs/0709.1027}{\texttt{arXiv:0709.1027}}.

\bibitem{Lai:2010vv}
H.-L. Lai\hrefCMSnoop {}{ {et~al.}, ``New parton distributions for collider
  physics'',} \textit{ Phys. Rev. D} \textbf{ 82} (2010) 074024,
  \href{http://dx.doi.org/10.1103/PhysRevD.82.074024}{\doi{10.1103/PhysRevD.82.074024}},
\href{http://www.arXiv.org/abs/1007.2241}{\texttt{arXiv:1007.2241}}.

\bibitem{PhysRevD.36.2019}
\hrefCMSnoop {}{T.~Sj{\"o}strand and M.~van Zijl, ``A multiple-interaction
  model for the event structure in hadron collisions'',} \textit{ Phys. Rev. D}
  \textbf{ 36} (1987) 2019,
  \href{http://dx.doi.org/10.1103/PhysRevD.36.2019}{\doi{10.1103/PhysRevD.36.2019}}.

\bibitem{Andersson:1998tv}
\hrefCMSnoop {}{B.~Andersson, ``{The Lund model}'',} \textit{ Camb. Monogr.
  Part. Phys. Nucl. Phys. Cosmol.} \textbf{ 7} (1997) 1,
\href{http://dx.doi.org/10.1016/0375-9474(87)90510-0}{\doi{10.1016/0375-9474(87)90510-0}}.

\bibitem{Alwall:2007fs}
J.~Alwall\hrefCMSnoop {}{ {et~al.}, ``{Comparative study of various algorithms
  for the merging of parton showers and matrix elements in hadronic
  collisions}'',} \textit{ Eur. Phys. J. C} \textbf{ 53} (2008) 473,
  \href{http://dx.doi.org/10.1140/epjc/s10052-007-0490-5}{\doi{10.1140/epjc/s10052-007-0490-5}},
  \href{http://www.arXiv.org/abs/0706.2569}{\texttt{arXiv:0706.2569}}.

\bibitem{bib:Nason:2004rx}
\hrefCMSnoop {}{P.~Nason, ``A new method for combining {NLO} {QCD} with shower
  {Monte} {Carlo} algorithms'',} \textit{ JHEP} \textbf{ 11} (2004) 040,
  \href{http://dx.doi.org/10.1088/1126-6708/2004/11/040}{\doi{10.1088/1126-6708/2004/11/040}},
\href{http://www.arXiv.org/abs/hep-ph/0409146}{\texttt{arXiv:hep-ph/0409146}}.

\bibitem{bib:Frixione:2007vw}
\hrefCMSnoop {}{S.~Frixione, P.~Nason, and C.~Oleari, ``{Matching NLO QCD
  computations with parton shower simulations: the POWHEG method}'',} \textit{
  JHEP} \textbf{ 11} (2007) 070,
  \href{http://dx.doi.org/10.1088/1126-6708/2007/11/070}{\doi{10.1088/1126-6708/2007/11/070}},
  \href{http://www.arXiv.org/abs/0709.2092}{\texttt{arXiv:0709.2092}}.

\bibitem{bib:Alioli:2010xd}
\hrefCMSnoop {}{S.~Alioli, P.~Nason, C.~Oleari, and E.~Re, ``{A general
  framework for implementing NLO calculations in shower Monte Carlo programs:
  the POWHEG BOX}'',} \textit{ JHEP} \textbf{ 06} (2010) 043,
  \href{http://dx.doi.org/10.1007/JHEP06(2010)043}{\doi{10.1007/JHEP06(2010)043}},
\href{http://www.arXiv.org/abs/1002.2581}{\texttt{arXiv:1002.2581}}.

\bibitem{bib:POWHEG_Dijet}
S.~Alioli\hrefCMSnoop {}{ {et~al.}, ``{Jet pair production in POWHEG}'',}
  \textit{ JHEP} \textbf{ 04} (2011) 081,
  \href{http://dx.doi.org/10.1007/JHEP04(2011)081}{\doi{10.1007/JHEP04(2011)081}},
  \href{http://www.arXiv.org/abs/1012.3380}{\texttt{arXiv:1012.3380}}.

\bibitem{Ball:2014uwa}
\hrefCMSnoop {}{{NNPDF} Collaboration, ``{Parton distributions for the LHC Run
  II}'',} \textit{ JHEP} \textbf{ 04} (2015) 040,
  \href{http://dx.doi.org/10.1007/JHEP04(2015)040}{\doi{10.1007/JHEP04(2015)040}},
\href{http://www.arXiv.org/abs/1410.8849}{\texttt{arXiv:1410.8849}}.

\bibitem{DAgostini:1994zf}
\hrefCMSnoop {}{G.~D'Agostini, ``A multidimensional unfolding method based on
  {Bayes'} theorem'',} \textit{ Nucl. Instrum. Meth. A} \textbf{ 362} (1995)
  487,
  \href{http://dx.doi.org/10.1016/0168-9002(95)00274-X}{\doi{10.1016/0168-9002(95)00274-X}}.

\bibitem{bib:RooUnfold}
\hrefCMSnoop {}{T.~Adye, ``Unfolding algorithms and tests using {RooUnfold}'',}
  in \textit{ {PHYSTAT} 2011 Workshop on Statistical Issues Related to
  Discovery Claims in Search Experiments and Unfolding}, H.~Prosper and
  L.~Lyons, eds., p.~313.
\newblock Geneva, Switzerland, 2011.
\newblock \href{http://www.arXiv.org/abs/1105.1160}{\texttt{arXiv:1105.1160}}.
\newblock
\href{http://dx.doi.org/10.5170/CERN-2011-006.313}{\doi{10.5170/CERN-2011-006.313}}.

\bibitem{Hocker:1995kb}
\hrefCMSnoop {}{A.~Hocker and V.~Kartvelishvili, ``{SVD approach to data
  unfolding}'',} \textit{ Nucl. Instrum. Meth. A} \textbf{ 372} (1996) 469,
  \href{http://dx.doi.org/10.1016/0168-9002(95)01478-0}{\doi{10.1016/0168-9002(95)01478-0}},
  \href{http://www.arXiv.org/abs/hep-ph/9509307}{\texttt{arXiv:hep-ph/9509307}}.

\bibitem{bib:jes8}
\hrefCMSnoop {}{{CMS Collaboration}, ``{Jet energy scale and resolution in the
  CMS experiment in pp collisions at 8 TeV}'',} \textit{ JINST} \textbf{ 12}
  (2017) P02014,
  \href{http://dx.doi.org/10.1088/1748-0221/12/02/P02014}{\doi{10.1088/1748-0221/12/02/P02014}},
\href{http://www.arXiv.org/abs/1607.03663}{\texttt{arXiv:1607.03663}}.

\bibitem{Chatrchyan:2012nj}
\hrefCMSnoop {}{{CMS Collaboration}, ``{Measurement of the inelastic
  proton-proton cross section at $\sqrt{s}=7$ TeV}'',} \textit{ Phys. Lett. B}
  \textbf{ 722} (2013) 5,
  \href{http://dx.doi.org/10.1016/j.physletb.2013.03.024}{\doi{10.1016/j.physletb.2013.03.024}},
  \href{http://www.arXiv.org/abs/1210.6718}{\texttt{arXiv:1210.6718}}.

\bibitem{hepdata}
\hrefCMSnoop {}{``{HEPD}ata record for this analysis'',} 2021.
\newblock
  \href{http://dx.doi.org/10.17182/hepdata.106642}{\doi{10.17182/hepdata.106642}}.

\bibitem{Butterworth:2015oua}
J.~Butterworth\hrefCMSnoop {}{ {et~al.}, ``{PDF4LHC recommendations for LHC Run
  II}'',} \textit{ J. Phys. G} \textbf{ 43} (2016) 023001,
  \href{http://dx.doi.org/10.1088/0954-3899/43/2/023001}{\doi{10.1088/0954-3899/43/2/023001}},
  \href{http://www.arXiv.org/abs/1510.03865}{\texttt{arXiv:1510.03865}}.

\end{thebibliography}\endgroup

\cleardoublepage \appendix\section{The CMS Collaboration \label{app:collab}}\begin{sloppypar}\hyphenpenalty=5000\widowpenalty=500\clubpenalty=5000\vskip\cmsinstskip
\textbf{Yerevan Physics Institute, Yerevan, Armenia}\\*[0pt]
A.M.~Sirunyan$^{\textrm{\dag}}$, A.~Tumasyan
\vskip\cmsinstskip
\textbf{Institut f\"{u}r Hochenergiephysik, Wien, Austria}\\*[0pt]
W.~Adam, T.~Bergauer, M.~Dragicevic, J.~Er\"{o}, A.~Escalante~Del~Valle, R.~Fr\"{u}hwirth\cmsAuthorMark{1}, M.~Jeitler\cmsAuthorMark{1}, N.~Krammer, L.~Lechner, D.~Liko, T.~Madlener, I.~Mikulec, F.M.~Pitters, N.~Rad, J.~Schieck\cmsAuthorMark{1}, R.~Sch\"{o}fbeck, M.~Spanring, S.~Templ, W.~Waltenberger, C.-E.~Wulz\cmsAuthorMark{1}, M.~Zarucki
% ...
\vskip\cmsinstskip
\dag: Deceased\\
1:  Also at Vienna University of Technology, Vienna, Austria\\
% ...
\end{sloppypar}
\end{document}